# Sprayable Thin and Robust Carbon Nanofiber Composite Coating for Extreme Jumping Dropwise Condensation Performance

*Matteo Donati[‡,1], Cheuk Wing Edmond Lam[‡,1], Athanasios Milionis[1], Chander Shekhar Sharma[1,2], Abinash Tripathy[1], Armend Zendeli[1], and Dimos Poulikakos[1,*]*

M. Donati[1], C. W. E. Lam[1], Dr. A. Milionis[1], Dr. C. S. Sharma[1,2], Dr. A. Tripathy[1], A. Zendeli[1], Prof. Dr. D. Poulikakos[1]
[1]Laboratory of Thermodynamics in Emerging Technologies, Department of Mechanical and Process Engineering, ETH Zurich, Sonneggstrasse 3, 8092 Zurich, Switzerland
[2]Thermofluidics Research Lab, Department of Mechanical Engineering, Indian Institute of Technology Ropar, Rupnagar, Punjab 140001, India

E-mail: dpoulikakos@ethz.ch

Condensation of water on metallic surfaces is critical for multiple energy conversion processes. Enhancement in condensation heat transfer efficiency often requires surface texturing and hydrophobicity, usually achieved through coatings, to maintain dropwise condensation. However, such surface treatments face conflicting challenges of minimal coating thermal resistance, enhanced coating durability and scalable fabrication. Here we present a thin (~ 2 μm) polytetrafluoroethylene – carbon nanofiber nanocomposite coating which meets these challenges and sustains coalescence-induced jumping droplet condensation for extended periods under highly demanding condensation conditions. Coating durability is achieved through improved substrate adhesion by depositing a sub-micron thick aluminum primer layer. Carbon nanofibers in a polytetrafluoroethylene matrix increase coating thermal conductivity and promote spontaneous surface nano-texturing to achieve superhydrophobicity for condensate microdroplets. The coating material can be deposited through direct spraying, ensuring economical scalability and versatility for a wide range of substrates. We know of no other coating for metallic surfaces that is able to sustain jumping dropwise condensation under shear of steam at 111 °C flowing at ~ 3 m s$^{-1}$ over the surface for 10 hours and dropwise condensation for an additional 50 hours. Up to ~ 900% improvement in condensation heat transfer coefficient is achieved compared to conventional filmwise condensation.

## 1. Introduction

Heterogeneous condensation on solid surfaces is an essential component of a wide range of industrial processes such as power plant condensers,[1] heat pipes in electronics cooling,[2] dew collectors in atmospheric water collection,[3] and in desalination or separation systems.[4] A significant amount of natural resources can be saved, if even a small enhancement in the overall efficiency of such processes can be achieved.[5] This in turn can ameliorate the continuous increase in energy demand, while mitigating greenhouse emissions and raw material consumption.[6]

One important component toward realizing aforementioned improvement consists of enhancing the heat transfer performance in condensers,[7] a key device in many industrial processes. Increased thermal efficiency of the condenser allows reduction in saturation pressure for steam condensation, for example, thereby increasing the enthalpy drop across the turbine and generating the same amount of electricity with less fuel consumption and carbon dioxide emissions.

It is well known that on hydrophobic surfaces, the vapor condenses in the form of discrete liquid droplets instead of a film. This effect is known as dropwise condensation (DWC) and can significantly enhance the heat transfer coefficient (HTC) as compared to filmwise condensation (FWC),[8] often occurring subsequently, if the condensate drops are not removed periodically under gravity to avoid formation of a continuous film.[1] Furthermore, superhydrophobic surfaces can promote an additional gravity-independent droplet departure mechanism through coalescence-induced droplet jumping that results in the ejection of much smaller droplets, leading to further heat transfer enhancement.[9,10] Compared to DWC, jumping dropwise condensation (JDWC) is known to further improve the HTC by up to ~ 30%,[11] and reduce the droplet departure size down to 500 nm.[12] Apart from these more common and rather broad condensate removal mechanisms, alternative passive pathways have been demonstrated, such as the cascading coalescence of condensed drops into microchannels,[13,14] lubricated surfaces,[15,16] and biphilic surfaces.[17]

Metallic surfaces in condensers, such as steel, aluminum or copper, are typically hydrophilic and DWC can be achieved on such surfaces by applying hydrophobic coatings. Numerous such coatings have been developed along with a range of coating techniques, including self-assembled monolayers (SAMs),[18,19] spraying,[20] physical (PVD)[21] and chemical vapor deposition (CVD).[22] However, these coatings need to address a number of conflicting challenges namely optimal wettability for efficient shedding of condensate, low thermal resistance, long-term mechanical durability, and scalability of the fabrication process. The coating materials are typically poor thermal conductors with thermal conductivities at the scale of ~ 0.1 – 0.5 W m$^{-1}$ K$^{-1}$,[23] significantly lower compared to metals. This imposes a restriction on the overall thickness of the coating, due to the associated proportional increase in the thermal resistance. Moreover, minimal thickness requirements also arise from the need to be conformal to micro- and nanotextures required for superhydrophobicity during condensation. Most of these surfaces have been tested under mild conditions (absence of shear flow at increased temperature) with few exceptions,[24,25] lacking indication of the durability over long timescales. Additionally, the simultaneous need for economical scalability remains unaddressed. Therefore, despite the existing promising concepts for improving condensation heat transfer, modern industrial condensation processes remain inefficient, and still rely on the use of uncoated metallic condenser surfaces which exhibit FWC.[26]

In this work, we demonstrate the material fabrication process and detailed characterization of condensation heat transfer and durability for a thin but robust superhydrophobic spray-coated polytetrafluoroethylene (PTFE) – carbon nanofiber (CNF) nanocomposite. Carbon-based materials have been shown to exhibit desirable characteristics in superhydrophobicity, robustness and electrical properties.[27–30] The addition of CNF in our nanocomposite not only leads to extreme water repellency during condensation due to spontaneous nanotexturing, but also creates a robust matrix with enhanced thermal conductivity.[31,32] Due to its unique surface topography and bulk composition, combined with its limited thickness achieved through optimization of the spray-coating process, this coating material is able to simultaneously address all the aforementioned challenges. In particular, we show JDWC to be the predominant mode of condensation for 10 hours, followed by DWC for an additional 50 hours, at the very challenging conditions of high steam temperature of 111 °C at 1.42 bar and simultaneous hydrodynamic abrasion from the exposure to shear stresses at ~ 3 m s$^{-1}$ steam flow. This performance is to the best of our knowledge unprecedented. Key factors to this enhanced robustness are the presence of the carbon nanofiber network[31,32] and the use of a submicron thick aluminum primer layer that improves adhesion to substrate without imposing any significant additional thermal resistance. We achieve a 9-fold enhancement in HTC compared to FWC on the reference uncoated substrate of the same material. All these outstanding performance characteristics are achieved through a facile, scalable and cost-effective fabrication process, rendering this coating process an excellent versatile and potential candidate for a host of substrates and applications.

## 2. Results and Discussion

### 2.1. Coating development

Our fabrication method consists of a number of simple steps, that can be easily scalable, while their combined result leads to the development of a multifunctional, i.e. hydrophobic, thermally conductive and robust, passive interface optimized for greatly improving condensation heat transfer. Furthermore, the combined cost of fabrication is maintained low and thus attractive for real-world industrial applications. We estimate the cost of our coating to be around USD 45 per $m^2$. For details, refer to Section S1 of the SI.

The first step of our process involves the application of a thin metal layer as primer. It is generally known that materials adhere preferentially to different substrates and this makes their applicability limited. Applying such an ultrathin sandwiched metal primer (between our coating and the thermally conductive substrate) alleviates this constraint. We illustrate this by successfully coating a copper substrate with PTFE after depositing a thin aluminum layer (~ 150 nm thick) by evaporation first that acts as a coating primer. The thermal resistance across the introduced aluminum-copper interface is negligible,[33] so is through the aluminum layer due to its very low thickness and high thermal conductivity. Furthermore, prior to the nanocomposite coating deposition, oxygen plasma is applied to the aluminum primer which helps clean and activate the surface with hydrophilic groups that enhance coating adhesion. These steps significantly improve the adhesion of the coating which otherwise has limited adhesion to copper.[34,35] Section S2 of the SI demonstrates the coating failure without a primer as the copper substrate is exposed after only 5 minutes of condensation at ~ 50 mbar.

This priming process is followed by a spray-coating approach and subsequent thermal annealing. We prepare a dichloromethane suspension of 1 µm PTFE microparticles and CNF of 20 – 200 µm length, which we spray directly on the Al-coated copper substrate. The thermal annealing is performed at 400 °C in a nitrogen-rich environment to inhibit the oxidation of the metal substrate. This melts the PTFE powder which then diffuses into the CNF network. In the resulting nanocomposite, 10 wt. % CNF is integrated within the PTFE matrix. For the optimization of the CNF concentration, refer to Section S3 of the SI. Each of these two components contributes to the desired functionalities of the coating.

PTFE, being a tough, hydrophobic and chemically resistant fluoropolymer, serves as a perfect base material as demonstrated by its already widespread commercial application as a water-repelling coating. On top of this, the nanofibers modify the surface and bulk properties to match our needs for condensation. The addition of CNF introduces micro- and nanostructures to the surface, required for superhydrophobicity. Regarding the bulk of the nanocomposite, the exceptional mechanical strength and thermal conductivity of CNF reinforces the PTFE matrix and lowers its thermal resistance respectively.[31,32] It is important to mention here that without the annealing step, the nanocomposite has a powdery, fragile morphology, thus this step is critical for achieving high durability. We term this nanocomposite coating as PTFE/CNF hereafter.

The coating thickness is reduced by tuning the total solid fraction of the spray suspension, while maintaining full substrate coverage with good uniformity and no surface defects as shown later in the SEM images in Section 2.2. This level of thickness is challenging to be achieved via spray coating,[36] especially with the addition of nanofibers in the sprayed dispersion that tend to expand the coating volume. We emphasize here that minimal coating thickness is a key aspect for achieving efficient condensation heat transfer, since a thick polymeric coating can counterbalance the heat transfer benefit gained from dropwise condensation on hydrophobic surfaces.[22] This is a subtle but important point in the design of heat-transfer-enhancing interface materials, since efforts towards minimizing the thickness of hydrophobic coating layers are typically accompanied with significant loss of long-term performance due to the lack of material durability.

We observe cross sections of the coating by the means of focused ion beam scanning electron microscopy (FIB SEM) of a pure PTFE and a PTFE/CNF coating that are both sprayed as 1 wt. % suspensions in dichloromethane. For further details on this characterization technique, refer to Section S4 of the SI. The determined coating thickness is 0.85 ± 0.30 μm for PTFE and 2.04 ± 0.35 μm for PTFE/CNF. The higher coating thickness of PTFE/CNF than that of PTFE with the same total used weight of materials can be attributed to the rather sparse packing of nanofibers. By weighing the surfaces before and after spraying, the area density of the PTFE/CNF coating is determined to be 7.0 ± 3.9 μg mm$^{-2}$, and for a coating thickness of 2 μm, the volume density is 3.5 ± 2.0 mg mm$^{-3}$. Thus, our careful spray optimization allows us to achieve a spray coating significantly thinner than conventionally expected.[36] We also find that this 2 μm-thick PTFE/CNF coating enables significant condensation heat transfer enhancement through JDWC while maintaining exceptional surface robustness as we show in the following sections. Other approaches to achieve thin hydrophobic coatings have either limited robustness (e.g. self-assembled monolayers) or unproven economical scalability.[22] An additional fundamental advantage of our approach compared to many existing ones is that with this method, no separate surface structuring is required (e.g. chemical etching or lithography), since all the hierarchical surface morphology required for superhydrophobicity is formed spontaneously upon spraying and subsequent annealing.

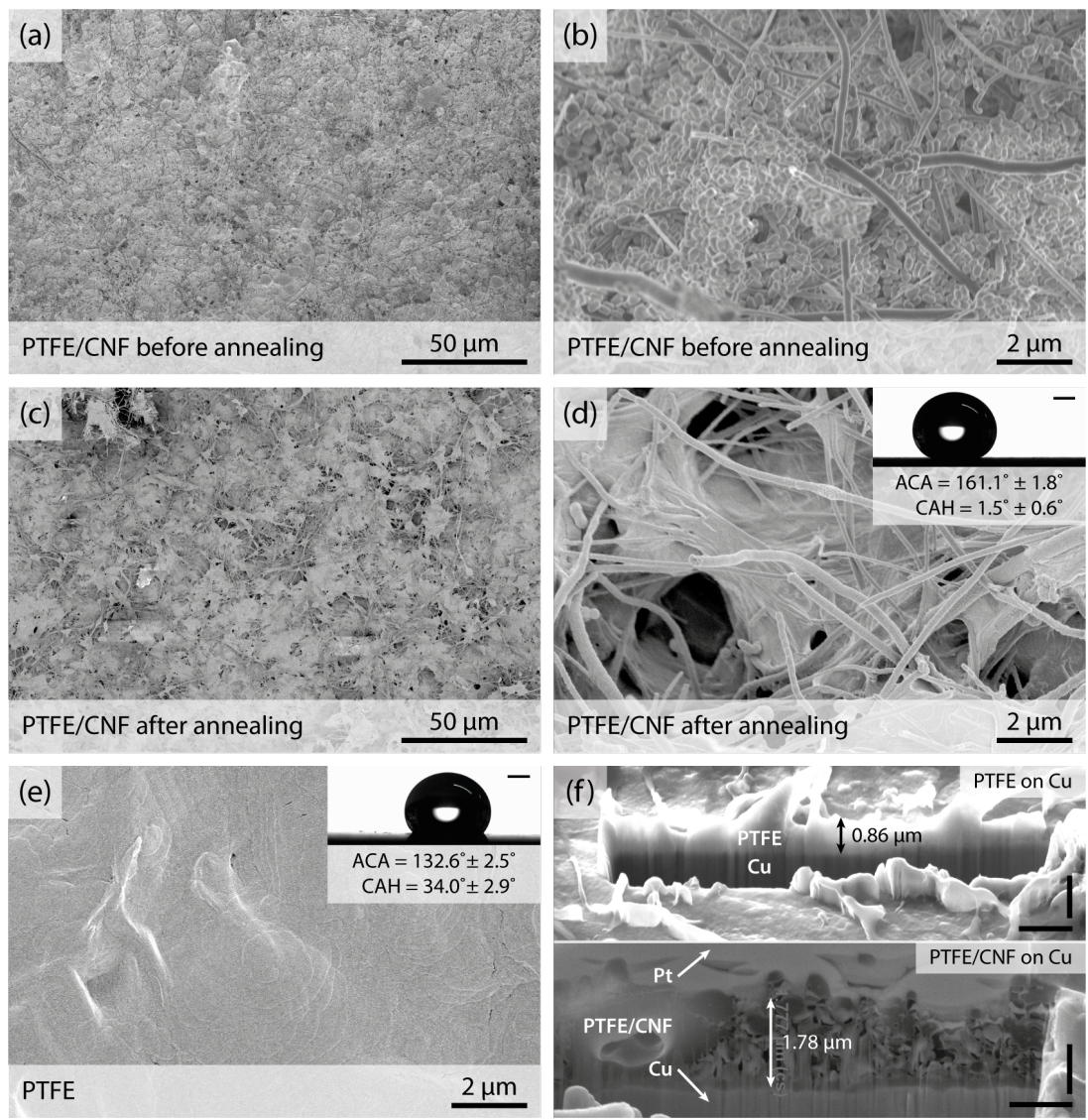

**Figure 1.** Scanning electron micrographs of the surfaces. PTFE/CNF (a, b) before and (c, d) after annealing; and (e) flat PTFE without CNF content. Insets: 10 µL droplets resting on the surfaces with the corresponding values of advancing contact angle (ACA) and contact angle hysteresis (CAH). Inset scale bar: 1 mm. (f) Cross sections of PTFE (top) and PTFE/CNF (bottom). Both coatings are sprayed on copper followed by annealing for thickness measurements. A layer of platinum is deposited in situ on top of the PTFE/CNF to allow easier determination of the top of the coating. Scale bar: 1 µm, vertical scale bars are shorter due to projection from an angled view.

## 2.2. Surface characterization

The images obtained with scanning electron microscopy (SEM) of the surface before and after annealing are shown respectively in **Figure 1**a, b and 1c, d. Prior to annealing, the as-sprayed PTFE/CNF is present in the form of discrete microparticles (PTFE powder) and fibers (CNF), with high porosity (Figure 1a, b). After annealing, the molten PTFE particles blend together and lock the CNF in place (Figure 1c, d). Still, after annealing, a high degree of hierarchical surface roughness (micro- and nanoscale) can be observed. In particular, the PTFE has formed microscopic flakes (Figure 1c) which enclose the high-aspect-ratio CNF. The nanofibers, randomly dispersed and bound by the polymer matrix, achieve the nanostructuring required to transition condensate microdroplets to mobile Cassie-like state.[37]

We choose two reference surfaces for our study, namely, 1) a CuO superhydrophilic surface,[38] to represent typical FWC on industrial condensers, and 2) a flat hydrophobic surface, to delineate the performance changes by the addition of CNF to PTFE. The superhydrophilic surface (Figure S4.1b of the SI) consists of ~ 1 µm-tall CuO nanoblades fabricated with a hot alkali process,[38] and it serves as a stable reproducible control case for FWC on copper surfaces. On the other hand, the flat hydrophobic surface is comprised of only PTFE (Figure 1e). Its fabrication process is similar to the PTFE/CNF nanocomposite, except that the CNF are excluded from the dichloromethane suspension. The overall solid weight fraction is kept as the same. There is no prominent roughness on the flat hydrophobic PTFE surface. For details on the fabrication process of all 3 surfaces, refer to the Experimental Section. Cross sections of the PTFE and PTFE/CNF coatings are shown in Figure 1f.

Measurements with contact angle goniometry confirm the superhydrophobic properties of our PTFE/CNF nanocomposite. Specifically, the advancing contact angle (ACA) is measured to be $161.1 \pm 1.8°$ while the contact angle hysteresis (CAH) is $1.5 \pm 0.6°$. These values show that the PTFE/CNF surface is significantly more water-repellent than PTFE alone without CNF, which exhibits an ACA of 132.6 ± 2.5º and a CAH of 34.0 ± 2.9º. The low water adhesion of PTFE/CNF is necessary for the phenomenon of coalescence-induced droplet jumping during condensation. Details on contact angle measurements are available in the Experimental Section.

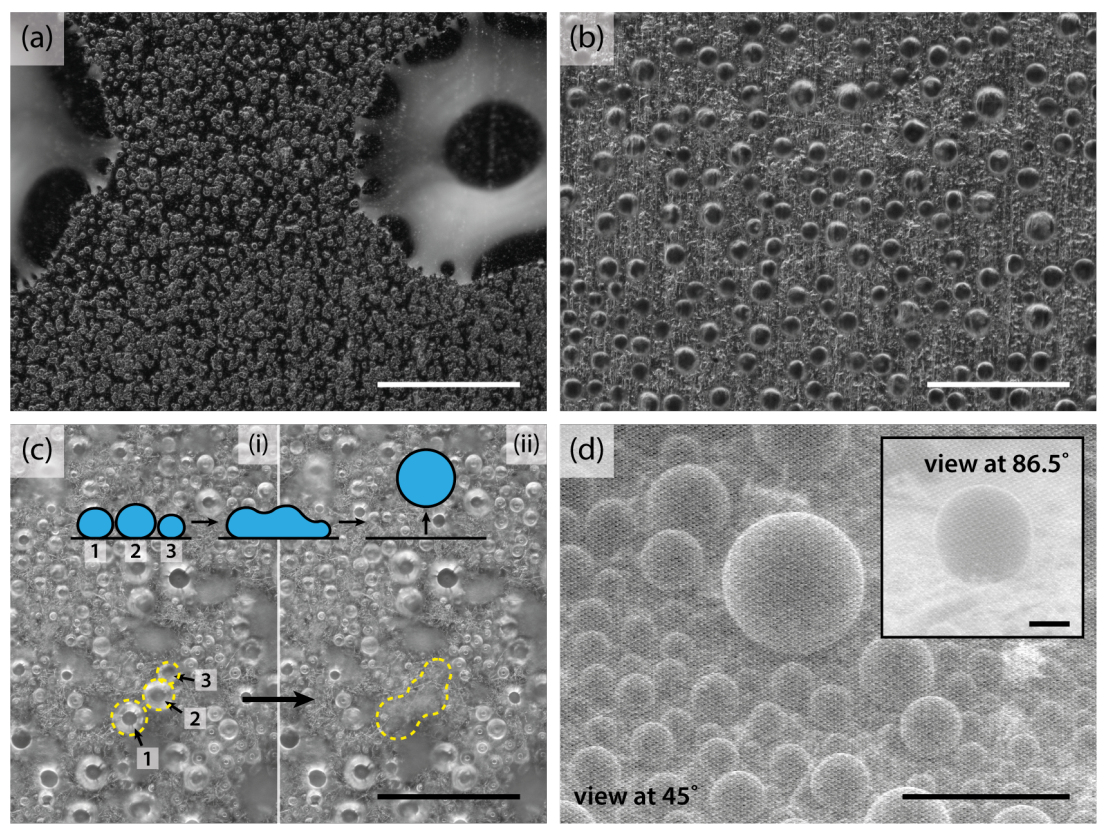

**Figure 2.** Optical micrographs showing the condensation behavior of the surfaces placed horizontally at room conditions, (a) superhydrophilic CuO; (b) hydrophobic PTFE; and (c) superhydrophobic PTFE/CNF, (i) droplets 1, 2 and 3 coalesce and leave the surface spontaneously, (ii) leaving the empty space enclosed by the dashed line. Schematic illustrates the droplet jumping process. (d) Condensation on PTFE/CNF under ESEM. Sample held at 45˚ with respect to the electron beam and maintained at ~ 2 °C with a cooling stage. Water vapor pressure is gradually increased to initiate nucleation. The nearly spherical shapes of condensate droplets suggest that the superhydrophobicity is maintained for condensed water and the surface is suitable for condensation heat transfer enhancement. Scale bar: 150 μm. Inset: Sample held at 86.5˚ during earlier stages of the condensation when the droplets are smaller. Inset scale bar: 10 μm.

## 2.3. Microscale condensation dynamics

To evaluate the microscale condensation behavior, we place our surfaces horizontally on a cooling stage and observe events with optical microscopy at room conditions. In **Figure 2**a, b, and c, we can see how condensation evolves on the superhydrophilic CuO reference surface, the PTFE flat reference surface and the PTFE/CNF surface respectively. A film of water is seen on the CuO surface in Figure 2a, and non-circular droplets with their contact lines pinned by the structures form on top of the film of water. On the other hand, on the flat PTFE surface in Figure 2b, distinct circular droplet contact lines are present, indicating that the droplets have shapes of spherical caps and are generally free from pinning by the surface, a key factor for droplet mobility to facilitate dropwise condensation. Coalescence-induced droplet jumping is observed only on the PTFE/CNF surface as shown in Figure 2c. 3 droplets (indicated by arrows in panel i) grow by condensation, coalesce and leave the surface spontaneously due to the conversion of the excess surface energy into upwards kinetic energy from a low-adhesion surface.[39] The empty space without water droplets, left behind after a coalescence jumping event (enclosed region in panel ii), enables new nucleation events to occur on the surface without undesirable thermal resistance from existing liquid water. As we will show afterwards, JDWC is the predominant departure mechanism for droplets condensed on PTFE/CNF. Details on these observations with optical microscopy is available in Section S5 of the SI.

We further observe the condensation behavior on PTFE/CNF in situ with an environmental scanning electron microscope (ESEM) (Video V1). As seen in Figure 2d, the condensed droplets exhibit nearly spherical shapes, suggesting high contact angles and a high repellency for small droplets during condensation, a condition not necessarily met on all superhydrophobic surfaces.[40] The inset in Figure 2d clearly shows the high contact angle for an individual condensate droplet. While a superhydrophobic surface may exhibit water repellency for deposited droplets, condensation nuclei may grow within the micro- and nanostructures to result in Wenzel-state droplets.[10,41,42] However, it is evident that the PTFE/CNF nanotexture is able to repel condensate droplets of the smallest size observable by our means of microscopy.[39] This is crucial to condensation heat transfer as discussed in the next section.

**2.4. Condensation heat transfer characterization**

A host of studies exist for novel micro- and nanostructures and coatings which facilitate DWC[43–47] and JDWC.[11,48,49] Yet, the vast majority is limited to surface wetting characterization, sometimes accompanied by measurements of the volume of condensed water collected from these surfaces.[46,47,49] However, these metrics do not necessarily directly correlate with heat transfer performance in terms of challenging but very valuable measurement of HTCs, which in the end is the decisive parameter for evaluating condenser efficiency. In other studies where the condensation HTCs are reported,[11,44,45,48] the testing conditions are often mild and far from a realistic industrial environment, where long-term performance is necessary. Here we investigate the heat transfer efficiency of PTFE/CNF coatings with direct measurements of HTCs for a range of steam temperatures and velocities.

We test the surfaces in 2 in-house steam flow experimental facilities, with the chamber of one designed to operate at a sub-atmospheric steam saturation pressure of 30 mbar, and the chamber of the other to operate at a steam saturation pressure above atmospheric of 1.42 bar. While the low-pressure chamber mimics realistic operating conditions of industrial condensers used in thermal power plants, the high-pressure chamber provides insights into the ability of PTFE/CNF to withstand more hostile conditions and provides an accelerated aging test for the coating, challenging its performance limits in conditions much harsher than those encountered in the majority of industrial settings. We believe that by evaluating the surfaces under varied environments, a comprehensive picture can be pieced together in terms of the applicability of the surfaces in different contexts.

*2.4.1. Heat transfer measurements at low pressure*

We first expose the surfaces to saturated steam at 30 mbar (corresponding saturation temperature 24.08 °C) in the low-pressure chamber. The surfaces are oriented vertically while the steam flows horizontally across at ~ 4 m s$^{-1}$ over the samples, which are placed on one end of a copper block, with its other end actively cooled with a recirculating chiller set at different temperatures. Condensation heat flux, steam temperature and sample surface temperature are measured as discussed with other experimental details in Section S7 of the SI.

The heat transfer performance of PTFE/CNF and the two reference surfaces at low steam saturation pressure are compared in **Figure 3**a and 3b.

In Figure 3a, the subcooling range decreases with increasing average HTCs of the surfaces, and there is a general shift of data points to lower subcooling values. This is a direct result of the surface temperature approaching the steam temperature due to the lower thermal resistance to condensation of a better performing surface. We average the HTCs obtained for each surface for comparison. At 30 mbar, the PTFE surface, which shows DWC, exhibits a 2.8x improvement in average HTC from the CuO surface which shows FWC, while PTFE/CNF which shows JDWC exhibits a 3.6x improvement. Figure 3b plots the heat fluxes of the 3 surfaces at different subcooling values, where the heat transfer improvement brought by PTFE/CNF can be seen from a different perspective. In general, for the same subcooling, PTFE/CNF permits a higher heat flux across the interface. Images of the surfaces during condensation are shown as insets in Figure 3b. The average droplet size on PTFE/CNF is visibly smaller than that on PTFE, a desirable indication of a more efficient removal of condensate droplets before they grow to significant larger sizes, critical for enhancing heat transfer.

This efficient condensate shedding for the case of PTFE/CNF is caused by frequent coalescence-induced droplet jumping from the surface as shown by the paths of the jumping droplets viewed from the side in Figure 3c and Video V2. While coalescence-induced droplet jumping is the predominant mechanism for droplet removal on PTFE/CNF, gravity-assisted droplet removal, which is also underpinned by the much larger departure diameters, is the sole mechanism on PTFE.

To quantify our observations, we measure the droplet diameters at their departure from PTFE/CNF and PTFE, as shown in Figure 3d. As droplet jumping departure is, in general, much more frequent than gravity-assisted departure, we measure departure diameters for all the droplets we can observe, i.e. above our resolution limit of ~ 20 μm per pixel, for ~ 2 s for PTFE/CNF and ~ 110 s for PTFE. We notice that most of the departing droplets on PTFE/CNF are restricted to diameters below 0.5 mm. We emphasize here that numerous jumping events may occur at diameters below our resolution limit. In general, the mean droplet departure diameter we can measure on PTFE/CNF ($\bar{d} = 290.2$ μm) is an order of magnitude smaller than on PTFE ($\bar{d} = 2564.4$ μm). The frequency of departure events is 95.3 cm$^{-2}$ s$^{-1}$ on PTFE/CNF whereas PTFE records a frequency of 1.1 cm$^{-2}$ s$^{-1}$, almost two orders of magnitude lower. Details on these measurements are described in Section S8 of the SI.

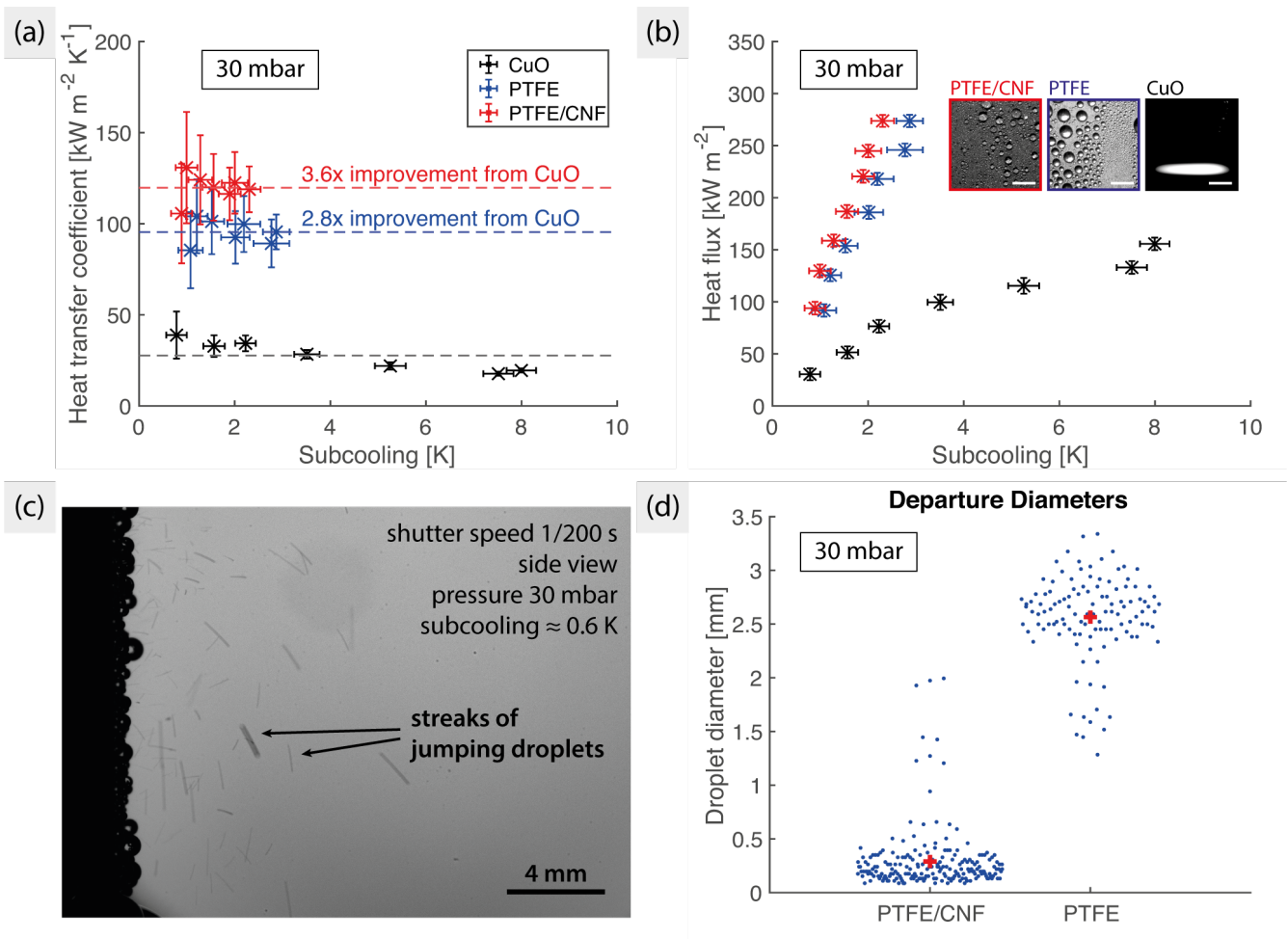

**Figure 3.** Measurements of vertically oriented superhydrophobic PTFE/CNF, hydrophobic PTFE, and superhydrophilic CuO under 30 mbar saturated steam flowing horizontally across over the surfaces. The flow speed is ~ 4 m s$^{-1}$ except in (c). (a) Heat transfer coefficients at different subcooling values. Dashed lines are averages of the measured HTCs for each surface. The data points shift to the lower subcooling regime for surfaces with higher HTC as the surface temperatures approach the steam temperature with lower thermal resistance to condensation at the interface. (b) Corresponding heat flux measurements of the 3 surfaces at different subcooling values under the same steam conditions. Insets are photographs of the 3 surfaces during condensation. The average droplet size on PTFE/CNF is visibly smaller than that on PTFE, and a film of water is observed on CuO. Inset scale bar: 2 mm. (c) Coalescence-induced jumping of condensate droplets from PTFE/CNF at a steam pressure of 30 mbar for an exposure time of 5 ms (shutter speed 1/200 s), viewed from the side. The paths of the jumping droplets are seen as streaks. These jumping events are highly frequent and comprise of droplets of various sizes. Steam flow speed is smaller than 4 m s$^{-1}$ here because of the modifications made to the flow chamber to allow side view. (d) Droplet departure diameters on PTFE/CNF and PTFE. Red crosses indicate mean measured diameters of a surface. Droplets jumping at smaller sizes may not be captured and measured when they are below the resolution limit of ~ 20 μm per pixel. Most of the departing droplets from PTFE/CNF have diameters below 0.5 mm whereas most from PTFE have diameters above 2.5 mm. The subcooling of both surfaces is 1 K.

*2.4.1. Heat transfer measurements at high temperature*

To evaluate the performance and resilience of the surfaces in more hostile environments, the 3 surfaces are subsequently placed vertically in the high-pressure flow chamber and exposed to superheated steam at 111 °C and 1.42 bar, flowing vertically downwards, i.e. in the gravitational direction.[25] We test the surfaces at two steam flow speeds, namely ~ 3 m s$^{-1}$ (laminar, Reynolds number Re ≈ 1300) and ~ 9 m s$^{-1}$ (turbulent, Re ≈ 3900), to study the effects of vapor shear on both droplet shedding and the coating. Re calculation is detailed in Section S10 of the SI. The principle of HTC determination is similar to that in the low-pressure flow chamber, as we describe the details of the setup in Section S9 of the SI.

The performance of the three surfaces under these condensation conditions are compared in **Figure 4**. At ~ 3 m s$^{-1}$, we observe a 5.6x increase in average HTC for DWC on the PTFE surface compared to FWC on the CuO surface, and a remarkable 9.2x increase for JDWC on the PTFE/CNF surface. The HTCs at ~ 9 m s$^{-1}$ are higher for all 3 surfaces, as expected, due to higher shear forces to shed non-jumping droplets on the surfaces and convective effects. The improvements of JDWC from FWC follow a similar trend, where we can observe a 4.2x increase in average HTC for DWC on PTFE and an 8.5x increase for JDWC on PTFE/CNF.

The above heat transfer characterization in the high-pressure chamber confirms that PTFE/CNF can deliver outstanding HTC improvement in significantly more demanding adverse environments as compared to the low-pressure mild conditions common in typical condensers. We subsequently test the surface performance under prolonged exposure to the high-temperature chamber conditions to characterize the surface robustness.

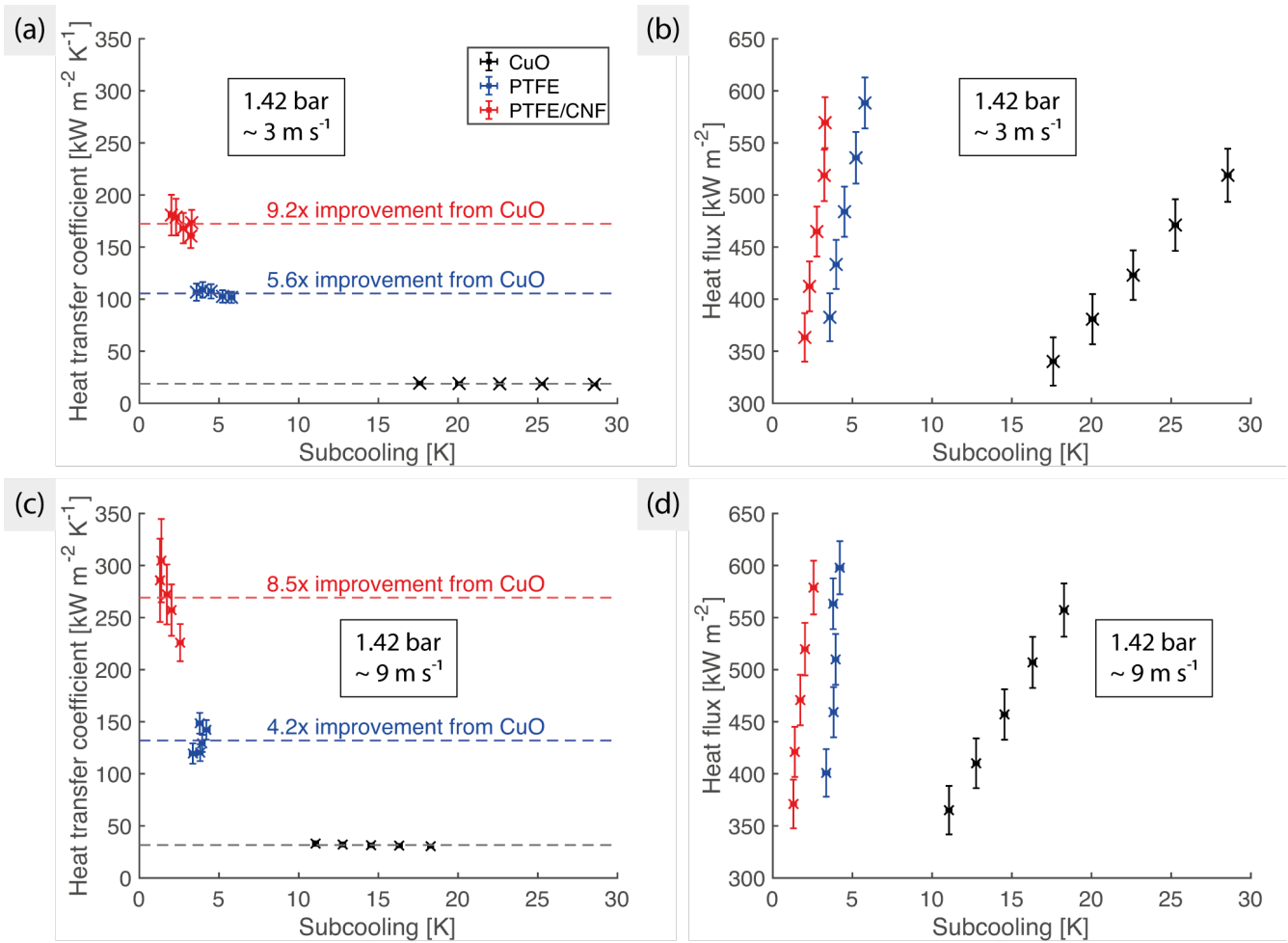

**Figure 4.** (a) Heat transfer coefficients of vertically placed superhydrophobic PTFE/CNF, hydrophobic PTFE and superhydrophilic CuO at different subcooling values under 111 °C superheated steam at 1.42 bar, flowing at ~ 3 m s$^{-1}$ downwards over the vertically placed surfaces. Dashed lines are averages of the measured HTCs for each surface. The data points shift to the lower subcooling regime for surfaces with higher HTC, similar to the low-pressure case. (b) Corresponding heat flux measurements of the 3 surfaces at different subcooling values at the same steam conditions. (c) Heat transfer coefficients of the 3 surfaces at different subcooling values under the same steam conditions except the flow speed is increased 3x, i.e. ~ 9 m s$^{-1}$, downwards. Similar shifting of the data points is observed. (d) Corresponding heat flux measurements of the 3 surfaces at different subcooling values at the same steam conditions, flowing at ~ 9 m s$^{-1}$.

## 2.5 Durability of PTFE/CNF

The main body of available research for the durability of surfaces for DWC and JDWC has mainly focused on testing them in flow-free environments at low steam temperatures, in which both shear and thermal stresses are absent. In studies of durable surfaces under more challenging conditions,[22,24,25,43] only DWC has been achieved. We test our nanocomposite in the high-pressure flow condensation chamber by exposing the surface to 1.42 bar superheated steam at 111 °C, flowing at ~ 3 m s$^{-1}$ for an uninterrupted period of 72 h. We track the changes in the condensation dynamics and the underlying surface degradation with a high-speed camera at several time intervals over the 72 hours.

Images of the PTFE/CNF surface are extracted at different times during the 72-hour durability test to show the evolution of condensation behavior over time (**Figure 5**a). At the very beginning of the experiment (0 h) the view is partially obstructed by unavoidable fogging on the viewing window due to condensate droplets jumping from the surface. At 2 h, a stable water film is formed on the viewing window, eliminating the fogging effect and establishing the needed transparency to view clearly the sample surface. The jumping behavior of the droplets from the surface continues uninterrupted. Up until 10 h, JDWC remains the predominant mode of condensation. After that, we observe an increase in the average droplet size on the surface and DWC takes over as the predominant mode. After 50 hours of further of continuous exposure, we finally observe a region of FWC emerging at 60 h at the bottom of the surface, enclosed by the red dashed line. This region expands slowly to take up a significant portion of the surface at 72 h, when we assume coating failure has occurred and we stop the experiment. The droplets in the DWC region have become less circular as well, indicating pinning of the three-phase contact lines by the underlying microstructure. However, despite the growth of this water film, we note that several droplet jumping events can still be observed close to the upper edge of the sample at 72 h.

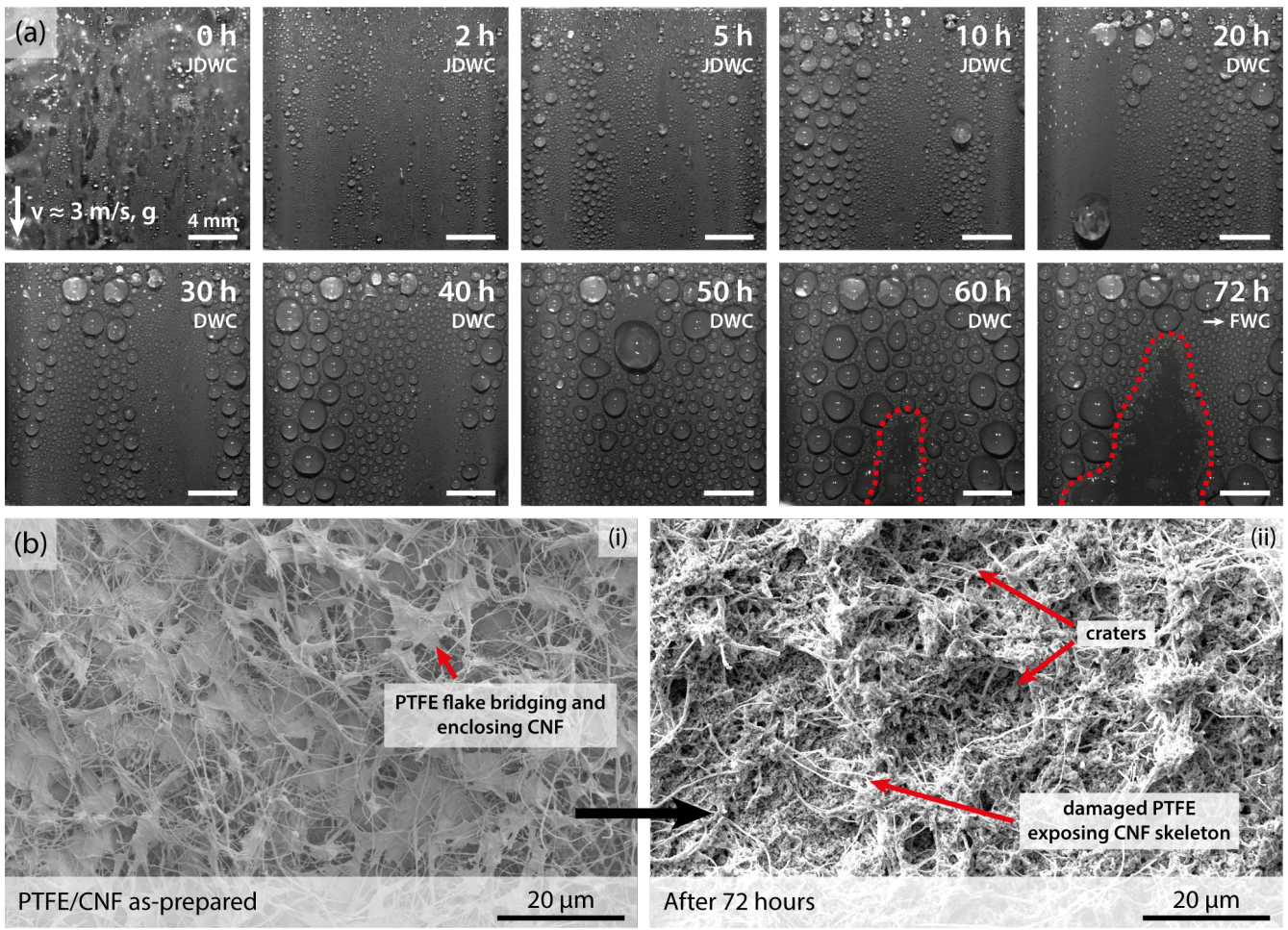

**Figure 5.** (a) Evolution of condensation behavior of PTFE/CNF in the high-pressure chamber over a period of 72 h. Superheated steam at 1.42 bar and 111 °C passes over the surface in the gravitational direction at ~ 3 m s$^{-1}$. FWC/ DWC/ JDWC: predominant mode of condensation. At the beginning of the experiment (0 h) the field of view is blocked by fogging, which is subsequently cleared as a smooth water film is formed on the viewport. Average droplet size on the surface grows larger as time progresses. At 60 h, a region of FWC emerges and grows, as enclosed by the red dashed lines. The droplets start to become less circular as they are pinned by the underlying microstructure. Coating failure is decided at 72 h when the FWC region has taken up a significant portion of the surface. (b) Scanning electron micrographs of (i) an as-prepared PTFE/CNF surface; and (ii) after the 72-hour durability test. The as-prepared PTFE/CNF exhibits PTFE flakes bridging across and enclosing CNF. These small and relatively flat PTFE regions are responsible for the microstructure and the securely embedded CNF is responsible for the nanostructure. After 72 h of steam exposure, the PTFE matrix is damaged and the CNF skeleton is exposed, leading to the loss of the original microstructure characterized by the PTFE flakes. A new microtopography is shaped in the form of craters, which are much larger than the flakes in (i). These craters trap condensation nuclei and facilitate Wenzel-state droplets and contact line pinning, leading to poor condensate removal and eventually FWC.

Apart from the long-term sustained JDWC and DWC demonstrated above, we attempt to quantify the strength of the coating by an estimation of the stresses it has tolerated. During the 72-hour durability test, the maximum shear stress exerted by the steam flow onto our surface is estimated, from flow calculations, to be at least 65 mPa (mean shear stress at least 57 mPa), which translates to an equivalent mass load of 6.6 µg mm$^{-2}$, 94% of the own mass of the PTFE/CNF coating (Section S10 of the SI). We note that additional shear stresses are present before the flow chamber is brought into steady operation, and the shear imposed by condensate droplets on the surface is not taken into account in our calculations, which consider only the steam flow. Refer to Section S10 of the SI for details. We suspect that as the average droplet diameter on the surface increases during the course of the experiment, the droplets exert a larger stress onto the coating thus accelerating further deterioration. Thus, the above stated shear stress tolerated by the coating is a conservative estimate. We observe that both the deterioration transitions from JDWC to DWC and from DWC to FWC start at the bottom of the surface. As non-jumping droplets depart by gravity, they sweep droplets along their path and reach the largest size at the bottom of the sample. Therefore, the shear stresses experienced at the bottom are constantly higher than the upper regions.

Finally, we compare the surface topography of PTFE/CNF before and after the durability test with SEM (Figure 5b), focusing on the region in which FWC becomes visible after 72 h of steam exposure. The topography of the surface is clearly modified. Despite the CNFs being still present and randomly dispersed over the surface, the PTFE matrix appears to be damaged and there are no longer smooth PTFE flakes bridging across and enclosing the fibers, leaving the skeleton of CNF exposed. The loss of PTFE creates new larger-sized and deeper microstructures in the form of craters on the surface. As nucleation occurs within these deeper craters, droplets may become trapped, resulting in the transition to the Wenzel-state. Eventually, the water condensed in the deeper and larger microstructures may interconnect and form a full film of water, much like the case of the superhydrophilic CuO nanostructured surface, thus leading to FWC.

The surface wettability of the PTFE/CNF surface after the durability test is also examined. We measure an ACA of 152.0 ± 1.9° and a CAH of 12.7 ± 2.9° at 3 random locations on the exposed condensing surface. The decrease in ACA and the increase in CAH is attributed to the change of the surface microtopography as explained above.

Our accelerated durability test proves that PTFE/CNF can withstand harsh conditions for an extended period of time while, in a host of practical applications, the conditions are much milder characterized by significantly lower steam temperature and pressures.[7,50]

## 3. Conclusion

We developed a versatile, scalable and economic method to fabricate exceptionally robust, sprayable superhydrophobic PTFE/CNF nanocomposite coating, able to sustain prolonged JDWC under extremely harsh conditions for 10 h and DWC for an additional 50 h. The key advantages of this method are: 1) no substrate pre-structuring is required, 2) appropriate surface priming with a metal layer allows for enhanced coating adhesion without compromising the heat transfer, 3) thin and superhydrophobic coating with thermally conductive nanofillers is realized via a facile spray method, and 4) high adhesion to substrate and the formation of the fiber network, together with PTFE annealing, give rise to an extremely robust coating given its minimal thickness (2 μm). In terms of heat transfer, this translates to an order of magnitude (9x) improvement in HTC compared to our FWC reference and almost doubled the HTC compared to the DWC reference. We can therefore conclude that we have carefully designed a passive and multifunctional, i.e. hydrophobic, thermally conductive and robust, material system that proves to be a promising superhydrophobic material for industrial condensing applications. Such a material could enable more efficient processes, for example, in steam power generation and other condensation-based energy conversion systems, thus contributing to the global reduction of carbon dioxide emissions.

## 4. Experimental Section

*Surface Fabrication*: Rectangular copper plates (EN CW004A, Cu ≥ 99.9%, Metall Service Menziken) precision cut to size 50 mm × 20 mm × 1.5 mm are used as the substrates for all the cases. For PTFE/CNF and PTFE surfaces, the copper substrates are first cleaned with acetone, isopropanol and hydrochloric acid (37%), sequentially, in an ultrasonicated water bath (USC300D, VWR). They are then rinsed thoroughly with deionized water and dried with nitrogen. A 150 nm thick layer of aluminum is then deposited onto the copper by e-beam evaporation (BAK 501LL, Evatec), followed by 3 minutes of oxygen plasma (Femto, Diener Electronic) at 100 W for surface activation. This thin layer is added to improve the coating adhesion to the substrate. To prepare the suspension for the PTFE/CNF coating, a vial of 90 mg PTFE (in powder form, 1 μm particle size, Sigma-Aldrich) is dispersed in 4.95 g of dichloromethane (Sigma-Aldrich) in an ultrasonicated water bath; and 10 mg CNF (> 98% carbon basis, 100 nm × 20 – 200 μm, Sigma-Aldrich) is mixed with 4.95 g of dichloromethane with a probe sonicator (Vibra-Cell VCX 130, Sonics) in another vial. The two vials are then mixed together, followed by an ultrasonicated water bath, resulting in a 1 wt. % PTFE/CNF suspension in dichloromethane, the suspension weighing a total of 10 g. This suspension is used to spray-coat 3 samples simultaneously with a VL double action – internal mix – siphon feed airbrush (Paasche Air Brush) at a distance of ~ 20 cm with an air pressure of 3 bar. To avoid precipitation, the suspension is continuously slightly shaken during spray. The surfaces are subsequently annealed in an atmosphere of nitrogen at 400 °C (FB1310M-33, Thermoline) for 30 minutes and cooled slowly to room temperature in the nitrogen environment. The preparation of the suspension for the PTFE coating is similar. 100 mg PTFE is mixed with 9.9 g dichloromethane in an ultrasonicated water bath. The suspension is then sprayed and annealed with the same parameters. For the CuO nanostructured surface, the same copper substrate is cleaned in an ultrasonicated water bath with acetone, isopropanol and deionized water, sequentially, then dried with nitrogen. The cleaned substrate is then immersed in 2 M hydrochloric acid for 20 – 30 seconds. Then it is transferred into a mixture of $NaClO_2$, NaOH and $Na_3PO_4:12H_2O$ at 95 °C ± 3 °C for 4 – 5 minutes, followed by rinsing with deionized water and drying with nitrogen.[38]

*Wettability Measurements*: Surface wettability is characterized by measuring advancing and receding contact angles with a goniometer (OCA 35, Dataphysics). For each surface, 5 measurements are taken at random locations.

*Observation of condensation with optical microscopy and ESEM*: Refer to Section S5 of the SI for details on the observation of condensation with optical microscopy and Section S6 of the SI for ESEM.

*Heat transfer characterization and droplet departure measurements*: Condensation heat transfer is characterized in 2 in-house flow chambers. The details on the structure of these chambers and their experimental procedures are available in Section S7 – S9 of the SI.

**Author Contributions**

‡ M.D. and C.W.E.L. contributed equally to this work. A.M., C.S.S., and D.P. conceived the research and provided scientific guidance in all of its aspects. M.D. and C.W.E.L. performed most of the experiments and processed the data. A.T. performed the FIB-SEM observations and measurements. A.Z. helped in the experimental work related to the coating durability evaluation. The manuscript was written through contribution of all authors.

**Acknowledgements**

We thank Peter Feusi and Jovo Vidic for their help with the experimental setup assembly as well as Mithulan Vasan for his assistance in the experiments. We also thank Daniel Notter, Yannick Zwirner and Jim Ferring for their contributions in experimental setup and protocol development. This project has received funding from the European Union's Horizon 2020 research and innovation program under grant number 801229 (HARMoNIC), and the Commission for Technology and Innovation, (CTI) under the Swiss Competence Centers for Energy Research (SCCER) program (Grant No. KTI.2014.0148).

**Sprayable Thin and Robust Carbon Nanofiber Composite Coating for Extreme Jumping Dropwise Condensation Performance**

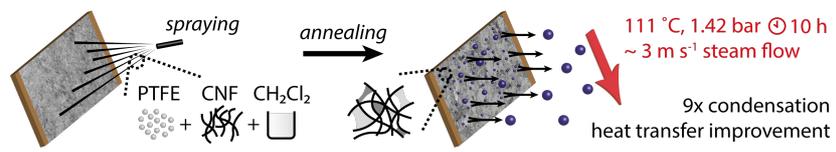

Supporting Information

**Sprayable Thin and Robust Carbon Nanofiber Composite Coating for Extreme Jumping Dropwise Condensation Performance**

*Matteo Donati, Cheuk Wing Edmond Lam, Athanasios Milionis, Chander Shekhar Sharma, Abinash Tripathy, Armend Zendeli, and Dimos Poulikakos\**

**Table of Contents**

# S1. Coating cost estimation

We estimate the cost of our coating at the industrial scale, based on the space available in our benchtop muffle furnace for annealing (0.01339 m$^2$). We base our calculations on this space because our furnace has the smallest area among all the equipment needed, i.e. one full fabrication procedure can therefore cover at most a size of 0.01339 m$^2$ with our available equipment. Also, our estimation is based on retail prices instead of potentially much lower wholesale prices for industrially large quantities.

Our estimation accounts for 1) components of the nanocomposite (dichloromethane, PTFE and CNF); 2) aluminum of the primer layer; and 3) the energy consumption for primer deposition, surface activation with plasma, probe sonication, ultrasonication, and annealing.

For this estimation we consider CNF from a supplier which offers a similar range of length (50 – 200 µm compared to 20 – 200 µm in the paper).

The cost of the materials is summarized in Table S1.1.

| Suspension component | Supplier | Amount [g] | Price [CHF] | Specific price [CHF/g] |
|---|---|---|---|---|
| Dichloromethane | Sidney Solvents | 266000 | 480 | 0.002 |
| PTFE | Nanografi | 1000 | 506.77 | 0.507 |
| CNF | American Elements | 1000 | 1146.62 | 1.147 |

Table S1.1: Cost of materials for the nanocomposite from selected suppliers

We compute the suspension material cost for a coating area of 0.01339 m$^2$, considering that 10 g of the suspension (9.9 g dichloromethane, 90 mg PTFE and 10 mg CNF) is needed to coat a surface of 0.003 m$^2$ (3 samples), as shown in Table S1.2.

| Suspension component | Amount [g] | Cost [CHF] |
|---|---|---|
| Dichloromethane | 44.187 | 0.080 |
| PTFE | 0.402 | 0.204 |
| CNF | 0.045 | 0.051 |
|  | **Total:** | **0.334** |

Table S1.2: Cost of materials to coat an area of 0.01339 m$^2$

The material cost of the aluminum primer is then considered. One evaporator run is needed to coat an area of 0.01339 m$^2$ and the 150 nm layer costs CHF 0.255.

We compute the total coating cost from the electricity prices of 3 countries, i.e. China, Switzerland, and the USA. The mean prices of electricity in 2019 for the business factor, based on a yearly 1,000,000 kWh consumption, are reported in Table S1.3.[1]

| Country | Price of electricity [CHF/kWh] |
|---|---|
| China | 0.090 |
| Switzerland | 0.152 |
| US | 0.110 |

Table S1.3: Electricity prices in China, Switzerland and the USA.

The energy consumption of the tools and their costs is thus listed in Table S1.4 for a coating area of 0.01339 m$^2$. For the electricity consumption of the furnace, we assume the power required to maintain the temperature at 400 °C during the 30-minute annealing process to be its maximum power of 1060 W.

| Tool | Power [kW] | Time [min] | Electricity consumed [kWh] | Cost (China) [CHF] | Cost (Switzerland) [CHF] | Cost (US) [CHF] |
|---|---|---|---|---|---|---|
| Evaporator | 1.6 | 12.5 | 0.333 | 0.03 | 0.051 | 0.037 |
| Plasma | 2.3 | 3 | 0.115 | 0.01 | 0.017 | 0.013 |
| Probe sonicator | 0.13 | 0.5 | 0.001 | 0.0001 | 0.00016 | 0.00012 |
| Ultrasonicator | 0.08 | 20 | 0.027 | 0.002 | 0.004 | 0.003 |
| Furnace | 1.06 | 30 | 0.53 | 0.048 | 0.081 | 0.058 |
| | | | **Total:** | **0.091** | **0.153** | **0.111** |

Table S1.4: Cost of electricity for equipment usage to coat an area of 0.01339 m$^2$, calculated from electricity prices in China, Switzerland and the USA.

The total cost to coat PTFE/CNF with electricity prices of the considered countries is summarized in Table S1.5, based on the area of 0.01339 m$^2$.

| Country | Cost of suspension components [CHF] | Cost of the aluminum of the primer [CHF] | Cost of electricity [CHF] | **Total cost [CHF]** |
|---|---|---|---|---|
| China | 0.334 | 0.255 | 0.091 | **0.680** |
| Switzerland | 0.334 | 0.255 | 0.153 | **0.742** |
| US | 0.334 | 0.255 | 0.111 | **0.700** |

Table S1.5: Total cost to coat an area of 0.01339 m$^2$ calculated from electricity prices in China, Switzerland and the USA

We note that at industrial scale, the prices further and significantly reduce as the sizes of equipment are optimized to make use of all available space (e.g. furnace and evaporator), as well as the ability to recycle dichloromethane from the nanocomposite after spraying. We report the cost of the coating per m$^2$ for the 3 countries taking the cost of dichloromethane into account and discounting it, as shown in Table S1.6.

| Country | Total coating cost w/ dichloromethane [CHF/m$^2$] | Total coating cost w/o dichloromethane [CHF/m$^2$] |
|---|---|---|
| China | 50.786 | 44.832 |
| US | 52.289 | 46.334 |
| Switzerland | 55.445 | 49.490 |

Table S1.6: Total cost to coat an area of 0.01339 m$^2$ calculated from electricity prices in China, Switzerland and the USA, taking the cost of dichloromethane into account and discounting it

Considering an approximately 1:1 exchange rate for CHF/USD, we conclude from this estimation that the cost of PTFE/CNF is ~ USD 45 per m$^2$.

## S2. Coating failure without the application of aluminum as primer

We demonstrate in this section the importance of applying aluminum as primer for the adhesion of PTFE/CNF to the substrate. The nanocomposite is applied directly on copper and the surface is then exposed to condensation under a low steam pressure of ~ 50 mbar and temperature of ~ 40 °C. After 5 minutes of exposure, part of the coating is visibly removed from the substrate, revealing the underlying copper (Figure S2.1a). At 73 minutes, condensate droplets appear at a larger diameter and deviate from the circular shape, which indicates pinning of contact lines (Figure S2.1b). Note that this experiment takes place at low pressure which is significantly milder than that in the high-pressure flow chamber.

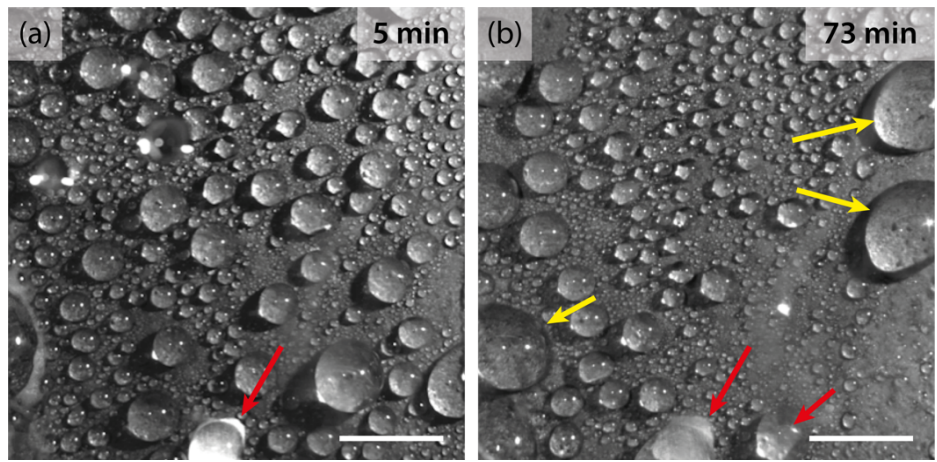

Figure S2.1: Condensation on PTFE/CNF without the aluminum primer, under a low steam pressure of ~ 50 mbar and temperature of ~ 40 °C. (a) After 5 minutes and (b) after 73 minutes of exposure. Red arrows are examples where the coating is visibly removed from the substrate. The copper substrate is revealed as more reflective brighter spots in the black-and-white images. Yellow arrows demonstrate some drops which deviate from circular shapes. Scale bar: 5 mm.

## S3. Optimization of CNF concentration in the suspension

Here we give an overview on how the concentration of CNF in the coating has been selected. We coat aluminum substrates with suspensions of PTFE/CNF in dichloromethane, keeping the total solid fraction of the suspension at 1 wt. % and varying the CNF content from 5 wt. % to 15 wt. %. The fabrication procedure is the same as described in the Experimental Section of the paper.

Surface wettability is then characterized by measuring the advancing and receding contact angles at 3 random locations. The samples are subsequently left for 30 minutes in hot deionized water at 95 °C, after which the surface wettability is characterized again. This test provides a first understanding of the coating durability in terms of maintaining its wettability under accelerated conditions, a crucial component to its long-term performance and ability to sustain jumping dropwise condensation as its predominant droplet removal mechanism.

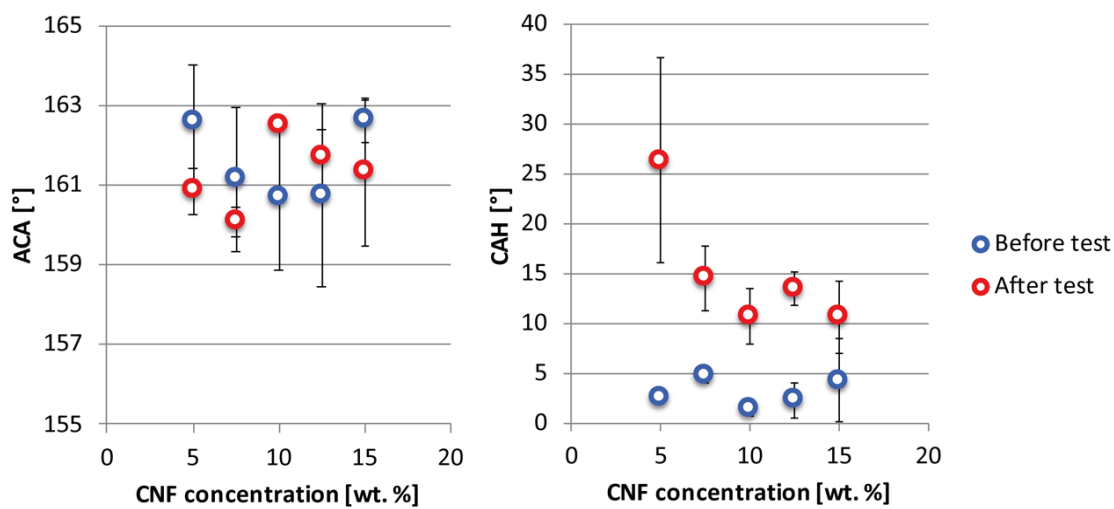

Figure S3.1: ACA and CAH on PTFE/CNF surfaces as a function of CNF concentration before and after their immersion in hot deionized water at 95 °C for 30 minutes.

ACA and CAH values before and after the test are reported as a function of the CNF concentration in Figure S3.1. Prior to the test, all surfaces are superhydrophobic, with mean ACAs above 160° and CAHs below 5°. After the immersion in hot water, whereas the ACAs remain almost unchanged for all the surfaces, the CAHs exhibit an increase. For 5 wt. % CNF, we observe that the CNF loading is not sufficient to maintain consistently low CAH after immersion in hot water, and therefore we proceed with the evaluation of increased CNF content. Among the remaining tested concentrations, 10 wt. % CNF gives the lowest CAH before and after the immersion test. We therefore select 10 wt. % as our optimal composition.

## S4. Coating thickness determination

We measure the height of the CuO nanoblades, and the thickness of our PTFE and PTFE/CNF coatings with focused ion beam (FIB) (Helios NanoLab 450S, FEI), as shown in Figure S4.1. The surfaces are tilted at $\alpha = 53°$ and a volume of 5 μm × 1 μm × 2.5 μm is drilled with the ion beam. Scanning electron microscopy is then used to image the cross sections. Cross sections of PTFE and PTFE/CNF are shown in Figure 1f of the paper. Figure S4.1b here shows a cross section of the CuO nanoblades. The height or thickness measurements are performed with the accompanying software taking the surface tilt into account. For each surface, ~ 45 measurements are taken at 5 different locations.

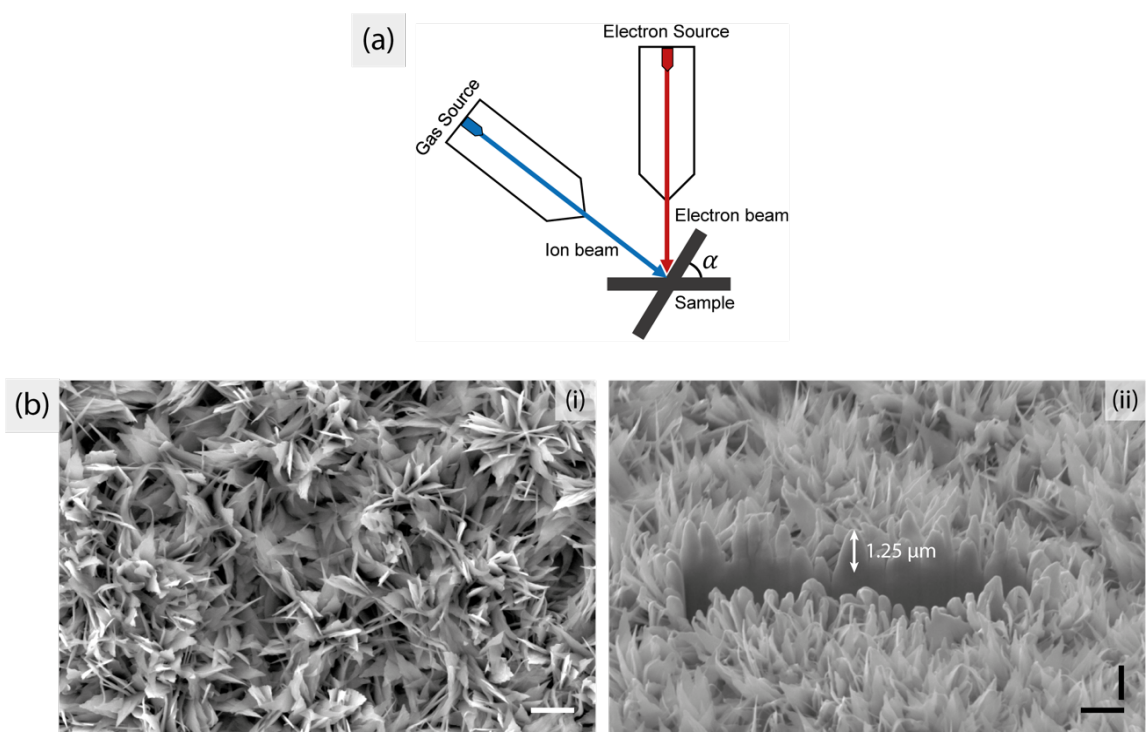

Figure S4.1: Determination of structure heights and coating thicknesses with focused ion beam (FIB). (a) An ion beam drills the surface tilted at $\alpha = 53°$ and the cross section is imaged by scanning electron microscopy (SEM). (b) Images obtained with SEM for the CuO surface. (i) Top view and (ii) cross section. Scale bar: 1 μm. Vertical scale bars are shorter because the surfaces are observed at an angle.

## S5. Observation of condensation with optical microscopy

Condensation on the 3 surfaces, i.e. CuO, PTFE, and PTFE/CNF, is observed under an optical microscope (BX60, Olympus) at room conditions (room temperature ~ 24 °C, relative humidity ~ 30%, dew point ~ 5 °C). The surfaces are placed horizontally on a stage, the temperature of which can be controlled by an electronic controller passing liquid nitrogen. Ambient temperature and relative humidity are measured to compute the dew point. Condensation takes place when the surface temperature is below the dew point.

Droplet jumping is observed only on the PTFE/CNF surface whereas the superhydrophilic CuO surface becomes quickly flooded with the condensate, and the PTFE surface shows dropwise condensation with no droplet jumping. Figure S5.1a depicts the formation of a condensate film within the structures of CuO and the growth of droplets on top of the film. While total wetting (static contact angle = 0˚) cannot be induced with only surface structures, these droplets are in the hydrophilic Cassie-Baxter state, and isolated "islands" of microstructures above the water film are observed.[2] The appearance of PTFE and PTFE/CNF surfaces just before condensation are shown in Figure S5.2.

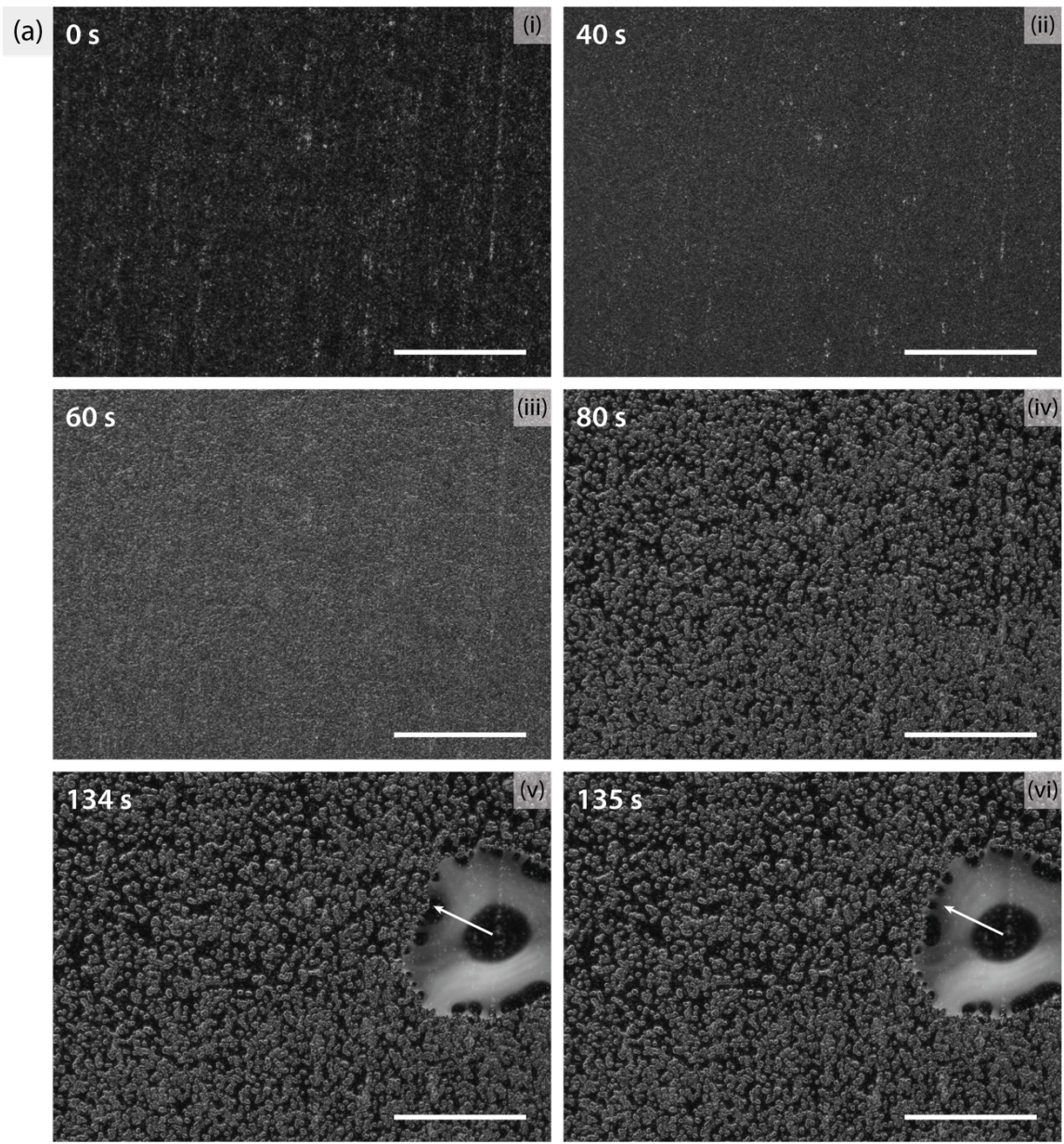

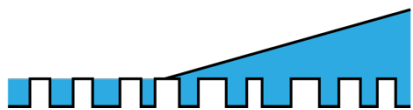

Figure S5.1: (a) Surface appearance of the CuO surface over time from (i) to (vi) as water condenses. A dry surface in black is depicted in (i). Nucleation is observed in (ii) and (iii) as the reflectivity of the surface changes with the formation of small water droplets. In (iv), a film of water has formed from the coalescence of these small droplets such that the surface reverts to its original reflectivity and the original black color is seen. Microstructures which are not submerged in this film are visible in a lighter color. (v) As more water is condensed, excess water forms a droplet as it rests in the hydrophilic Cassie-Baxter state on top of the water film. The white arrow indicates pinning of the contact line of the droplet by the microstructures. (vi) The white arrow points at the same location as in (v). The contact line has overcome the pinning and moved forward as the droplet grows by condensation. Scale bar: 150 μm. (b) Schematic of the contact line of a droplet in the above explained hydrophilic Cassie-Baxter state. The droplet rests on effectively a solid/liquid composite.[2]

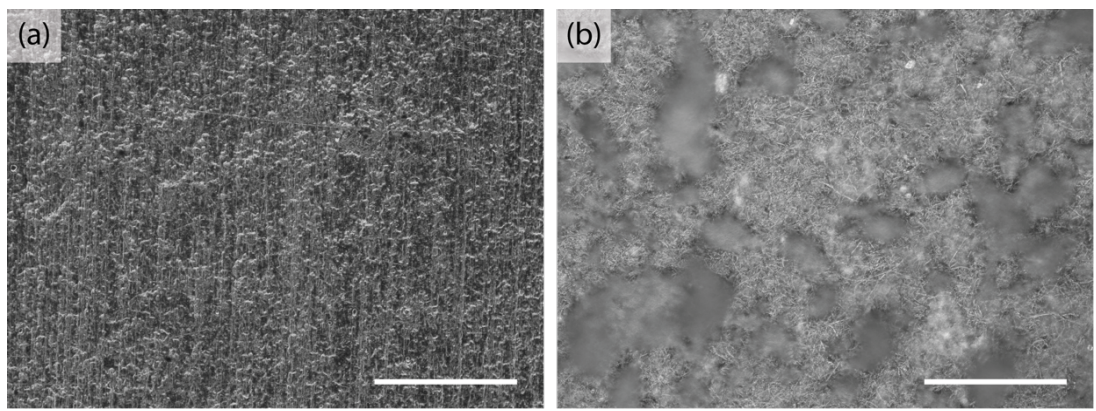

Figure S5.2: (a) PTFE and (b) PTFE/CNF at approximately the same locations as Figure 2 in the paper just before condensation. Scale bar: 150 μm.

**S6. Observation of condensation with environmental scanning electron microscopy**

A PTFE/CNF surface is placed in an environmental scanning electron microscope (ESEM) (Quanta 600, FEI) on a custom-made copper platform which sets the sample at an angle with respect to the electron beam. The platform is cooled with a recirculating chiller and is maintained at ~ 2 °C. Pure water vapor is introduced into the chamber after evacuation. Condensation begins to take place after the water vapor pressure exceeds the saturation pressure corresponding to the surface temperature. ESEM allows an angled observation without significant loss of depth of field, confirming the high contact angle of droplets of PTFE/CNF during condensation.

## S7. Heat transfer measurements at low pressure

### (a) Overall experimental setup

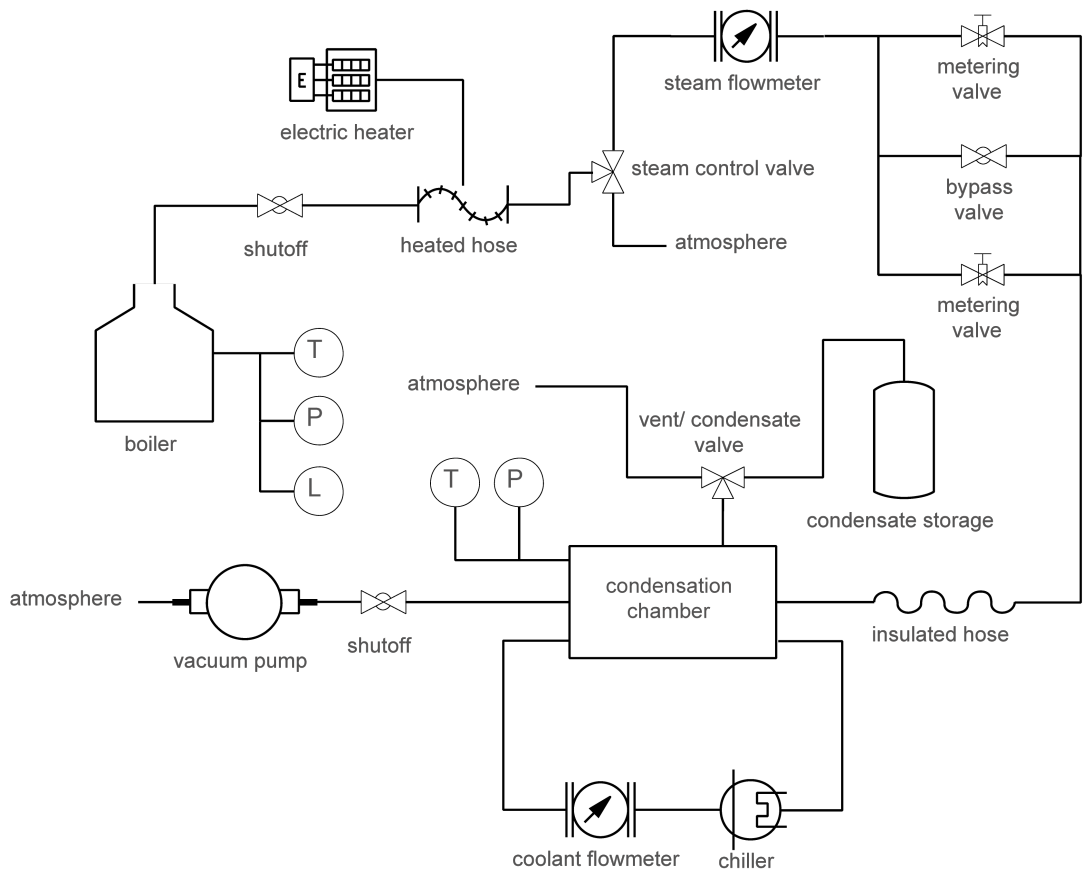

Figure S7.1: Schematic of the overall experimental setup for low-pressure condensation

This section describes the low-pressure flow chamber and experimental procedures. A schematic of the overall setup is shown in Figure S7.1.

A thermally insulated boiler is used to boil deionized water to steam. The temperature, pressure and liquid level of the water inside the boiler are constantly monitored. Water from the boiler passes through an electrically heated hose to prevent condensation within the hose. A 3-way steam control valve is used to switch the steam supply between the chamber and atmosphere. The flowmeter (FAM3255, ABB), installed later into the setup, measures the mass flow rate of the steam and thus provides a steam flow speed estimate of ~ 4 m/s over the sample surface in the chamber, after dividing the steam volume flow rate by the area available to the flow. The bypass valve enables evacuation of non-condensable gases up to the 3-way steam control valve. 2 metering valves are arranged in parallel for coarse and fine control of the steam pressure in the chamber. A condensate storage is connected to the chamber through a 3-way vent/ condensate valve. The same valve is used to vent the chamber after experiments. A recirculating chiller (WKL 2200, LAUDA) cools the copper cooler located at the back of the test sample and its flow rate is monitored with a flowmeter (SITRANS FM MAG5000 and SITRANS FM MAG 1100, SIEMENS). During the experiment, the vacuum pump (RC 6, VACUUBRAND) maintains a stable flow of saturated steam over the surface.

The following protocol is adopted for the experiments to ensure minimum non-condensable gases in the flow chamber and precise measurements of heat transfer coefficients. Before introducing steam into the chamber, the 3-way steam control valve, after the hose, is opened to atmosphere and the water is continuously boiled at 1.4 bar for 30 minutes to degas and to expel non-condensable gases from the boiler. At the same time, the vacuum pump evacuates the condensation chamber to < 0.01 mbar (1 Pa), the lower limit of our measuring capability, to remove non-condensable gases from within the chamber and the connecting tubes and hoses up to the 3-way steam control valve. After 30 minutes of degassing, the bypass valve is open and closed repeatedly for the pump to remove residual gases trapped inside the valve. The bypass valve then remains closed. The chiller is set to an initial coolant temperature of 20 °C to avoid excessive subcooling as fresh steam reaches the condensing surface. Steam is then introduced into the chamber by turning the steam control valve. Steam expands after the two metering valves, which control the chamber pressure. The low-pressure steam passes through an insulated hose to prevent condensation before entering the chamber. Pressure and temperature sensors are connected to the chamber to monitor steam and surface conditions. As soon as steam is introduced, the coolant temperature is reduced to different values to have increasing subcooling of the surface. Once a value is reached, the coolant flow rate is set to 180 L/h ± 10 L/h. Valves are adjusted to stabilize the system. When the system becomes stable, steady state is reached and 120 readings over 1 minute are recorded, during which no adjustments are made to the setup. This concludes the heat transfer measurement at 1 subcooling value and then the next coolant temperature is set. The procedure is then repeated to obtain measurements for a range of subcooling values.

The inner structure of the chamber is described in the following section.

**(b) Low-pressure condensation flow chamber**

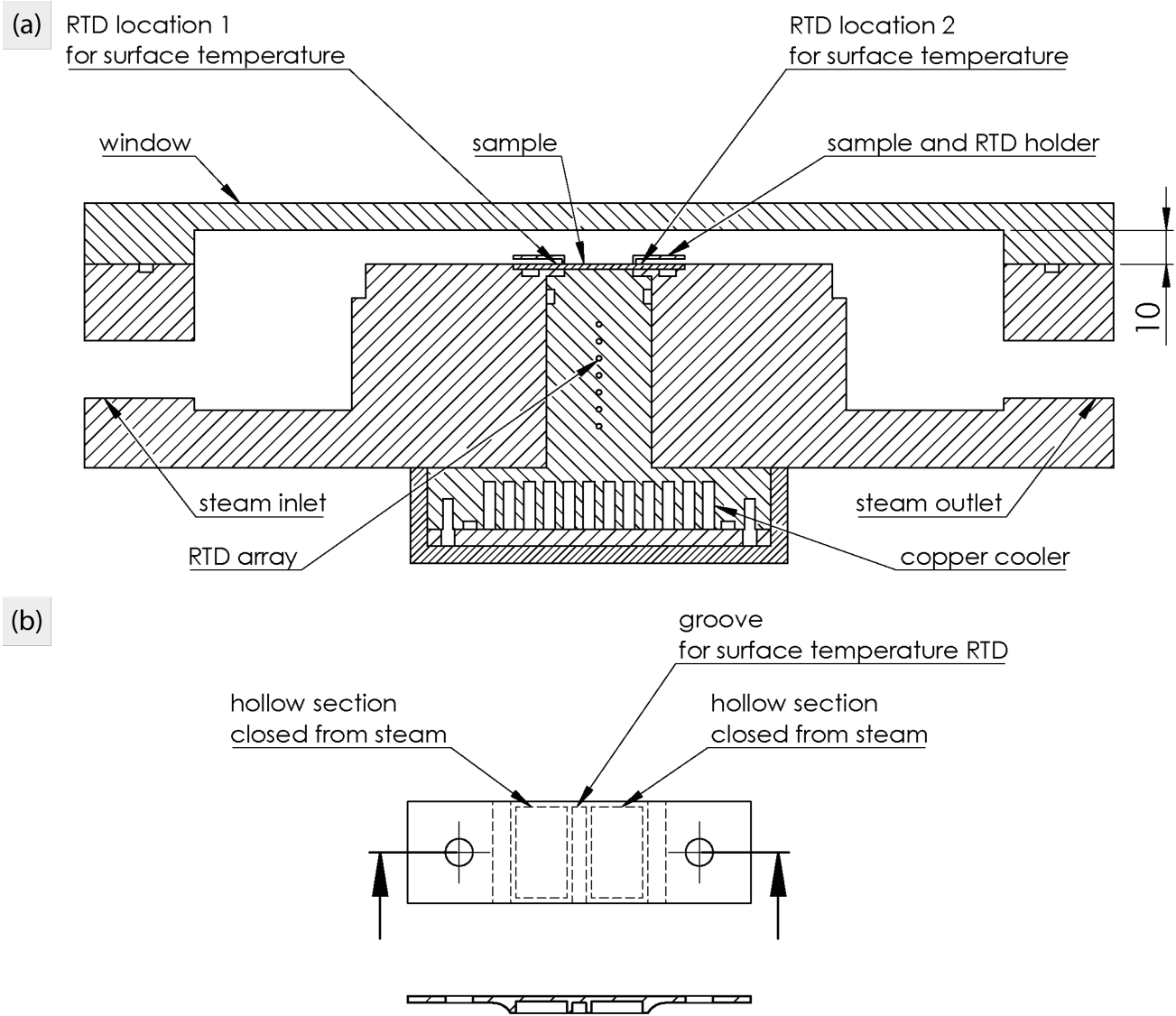

Figure S7.2: Schematics of the low-pressure condensation chamber. (a) Top view cross-section. (b) Sample and surface temperature RTD holder

Figure S7.2a presents a top view cross-section schematic of the condensation chamber. Saturated steam is introduced into the chamber from the left. The steam then passes into the channel of 10 mm height over the sample at ~ 4 m/s. At the back of the sample, a copper block with a thermal conductivity of 394 W/mK (CW004A, Durofer AG) is installed to extract heat from the steam as it condenses on the surface. A thin layer of thermal paste is added between the sample and the copper block (not explicitly drawn). The vacuum pump is connected to the steam outlet of the chamber. Along the axis of the cylindrical section of the copper cooler, 7 Class A Pt100 resistance temperature detectors (RTDs) are placed at a precise pitch of 5 mm to measure the temperature gradient and hence condensation heat flux. The sides of the copper where the RTD array is located are insulated with an air gap (not explicitly drawn) and polyether ether ketone (PEEK, Amsler & Frey AG), with a thermal conductivity of 0.25 W/mK, to facilitate one-dimensional heat conduction along the array. At the back of the cooler, a coolant is recirculated using the chiller.

Chamber pressure is measured by two pressure sensors (CMR 361 and CMR 362, Pfeiffer Vacuum) which together measure a total range from 0.01 mbar to atmospheric. Close to each pressure sensor an RTD is placed to confirm steam saturation conditions. For the surface temperature, 2 RTDs are taped and sealed with Kapton on the sides of the sample so there is no direct contact with incoming steam. The surface temperature RTDs are kept as close to the central 20 mm × 20 mm condensing surface as possible, meanwhile covered with hollow 3D-printed polycarbonate mounts (Figure S7.2b) so that the incoming steam does not directly come into contact with the 20 mm × 15 mm sides of the sample (sample total size 50 mm × 20 mm, condensing surface 20 mm × 20 mm), and due to their hollow structure, the top surface of the mount is thermally insulated from the surface of the sample, limiting the effect of extra heat transfer area from the mounts.

All sensors are connected to a data acquisition system and readings are recorded every 500 ms ± 0.6 ms.

**(c) Computation of heat transfer coefficients**

The heat flow $q$ passing through the cooler is computed using the temperature gradient measured with the RTD array:

$$q = k_c A_c \frac{dT}{dx}$$

where $k_c$ is the thermal conductivity of the cooler, $A_c = 735$ mm² is the cross-sectional area of the cooler, and $dT/dx$ is the constant thermal gradient along the array. We base our heat flux $q''$ on the area of the exposed condensing surface $A_e = 400$ mm², considering that the conservation of energy requires the heat flow in the cooler along the array to be equal to the heat flux across the condensing surface:

$$q'' = \frac{q}{A_e}$$

The heat transfer coefficient (HTC) $h$ is then computed:

$$h = \frac{q''}{\Delta T}$$

where $\Delta T = T_{\text{inf}} - T_s$ is the subcooling, defined as the difference between the steam temperature $T_{\text{inf}}$ and the surface temperature $T_s$ respectively.

The steam temperature $T_{\text{inf}}$ is determined from the mean of 2 RTDs on each side of the chamber away from the condensing surface. Similarly, the sample temperature $T_s$ is determined from the 2 RTDs on each side of the sample.

$dT/dx$ is computed from a least-square linear fit of the temperatures measured with the RTD array.

Before taking the value for each data point, steady state is confirmed over a minute of readings at 2 Hz. The mean of these 120 readings is reported in the main text.

### (d) Uncertainty propagation

Uncertainty propagation is computed for our measurements.[3] Uncertainties originating from the RTDs for steam temperature, surface temperature, and the temperature gradient in the cooler are taken into account and propagated to the uncertainties in subcooling, heat flux, and HTC values.

(i) Uncertainty in subcooling

Subcooling is defined as the difference between the steam temperature and the surface temperature. All RTDs employed in the low-pressure flow chamber are of Class A, with an uncertainty of $\pm(0.15 + 0.002T)\,°C$, where $T$ is the measured temperature. For the temperatures we consider in this study, we simplify the expression by assuming an uncertainty of $\delta_{\mathrm{RTD}} = \pm 0.2\,°C$ for 1 RTD in subsequent calculations.

The uncertainty in steam and surface temperatures is the same as one RTD:

$$\delta_{T_{\mathrm{inf}}} = \delta_{T_{\mathrm{s}}} = \delta_{\mathrm{RTD}}$$

The uncertainty in subcooling is then obtained:

$$\delta_{\Delta T} = \sqrt{\delta_{T_{\mathrm{inf}}}^2 + \delta_{T_{\mathrm{s}}}^2}$$

(ii) Uncertainty in heat flux

The uncertainty in heat flux results from the uncertainty in the determination of the slope of the temperature gradient. The uncertainty in the linear fit $\delta_{\text{lin\_fit}}$ is calculated as follows:

$$\delta_{\text{lin\_fit}} = \delta_{\text{RTD}} \sqrt{\frac{N_{\text{RTD}}}{N_{\text{RTD}} \sum x_{\text{RTD}}^2 - (\sum x_{\text{RTD}})^2}}$$

where $N_{\text{RTD}} = 7$ is the total number of RTDs and $x_{\text{RTD}}$ is the location of the RTDs in the array.

The uncertainty in the linear fit is propagated to the uncertainty in heat flux, taking into consideration the thermal conductivity of the cooler, and the difference in area between the cross-section of the cooler and the condensing surface.

$$\delta_{q''} = k \frac{A_c}{A_e} \delta_{\text{lin\_fit}}$$

(iii) Uncertainty in HTC

HTC is obtained from the division of heat flux by subcooling. The uncertainty is propagated as follows:

$$\delta_h = \sqrt{\left[\frac{1}{(T_{\text{inf}} - T_s)} \delta_{q''}\right]^2 + \left[\frac{-q''}{(T_{\text{inf}} - T_s)^2} \delta_{T_{\text{inf}}}\right]^2 + \left[\frac{q''}{(T_{\text{inf}} - T_s)^2} \delta_{T_s}\right]^2}$$

$$\delta_h = \sqrt{\left[\frac{1}{(\Delta T)} \delta_{q''}\right]^2 + \left[\frac{-q''}{(\Delta T)^2} \delta_{T_{\text{inf}}}\right]^2 + \left[\frac{q''}{(\Delta T)^2} \delta_{T_s}\right]^2}$$

Here we take 120-reading mean values of $T_{\text{inf}}$, $T_s$ and $q''$ to compute the uncertainty of the 120-reading mean value of $h$.

(iv) Overall uncertainty

The uncertainties in subcooling, heat flux, and HTC above arise from the random errors of the sensors themselves. The reported values in the main text are mean values of the 120 readings in one minute. To provide an indication of the uncertainty from the variations over time, the standard deviations of the 120 readings are added to the corresponding uncertainty. Thus, the resulting uncertainty in subcooling, heat flux, and HTC are $\delta_{\Delta T} + \sigma_{\Delta T}$, $\delta_{q''} + \sigma_{q''}$, and $\delta_h + \sigma_h$ respectively, where $\sigma$'s refer to the standard deviation of the corresponding value over 120 measurements in one minute. We report these overall uncertainties in the main text for the low-pressure flow chamber.

## S8. Observation of condensation at low pressure

### (a) Modifications to chamber window for observation

The low-pressure flow chamber is modified in different ways to enable observation from the front and the side.

(i) Front view

As the steam temperature (~ 24 °C) is slightly higher than ambient temperature (~ 20 °C), fogging occurs on the inside surfaces of the chamber. For observation from the front (insets in Figure 3b, and departure diameters of Figure 3d), a window with a heated viewport is used. The cross-sectional area of the flow channel remains similar so that the steam flow speed is maintained at ~ 4 m/s. A halogen lamp is used as illumination.

(ii) Side view

For side observation (photograph of Figure 3c), due to the limitation of space from the side of the original window (Figure S7.2a), a window with a higher height is used and the steam speed in Figure 3c is smaller than 4 m/s due to a larger cross-sectional area of the flow channel. Similarly, this window has a heated viewport on the side to avoid fogging. Illumination is provided with LED on the other side of the window as backlight. To enhance the visibility of droplets jumping from the surface, minor digital contrast enhancements are applied to the photograph of Figure 3c (Camera Raw Filter in Photoshop, Adobe) and Video V2 (Premiere Pro, Adobe).

These windows with heated viewports are only used in observation experiments. For heat transfer measurements, windows without heating are used to eliminate heating effects on temperature measurements in the chamber.

**(b) Measurement of droplet departure events**

Droplet departure diameters are measured with the front-view heated window as described above. To eliminate boundary effects from the surface edges and sample mounts, the captured videos are cropped to an area of 10 mm × 10 mm at the center of the condensing surface. ImageJ is used to obtain the diameters with a manual circular fit of the droplets in concern.

Due to the vast difference in the frequency of departing droplets between the PTFE/CNF and PTFE surfaces, ~ 2 s and ~ 110 s of footage are analyzed respectively. Footage for PTFE/CNF are recorded at 5000 fps and that for PTFE are at 50 fps, so to allow longer recording times required for the longer timescales of gravity-assisted droplet departure on the PTFE surface. Shutter speed is set at a constant 1/5000 s for all footage for consistency as well as to reduce motion blur. The resolution of these footage is ~ 20 μm per pixel.

For jumping events, which are exclusive to PTFE/CNF, diameters are measured just after jumping. For gravity-assisted droplet removal which may occur on both PTFE/CNF and PTFE, diameters are measured just before the droplets leave the observation area at its lower boundary. Departure diameters are reported in the main text as beeswarm plots (Figure 3d) from an existing MATLAB code.[4]

Departure frequency is obtained from the division of the number of events on each surface by the footage time. For PTFE/CNF, the frequency reported is likely an underestimation as some of jumping events may fall below our resolution limit.

## S9. Heat transfer measurements at high pressure

### (a) Overall experimental setup

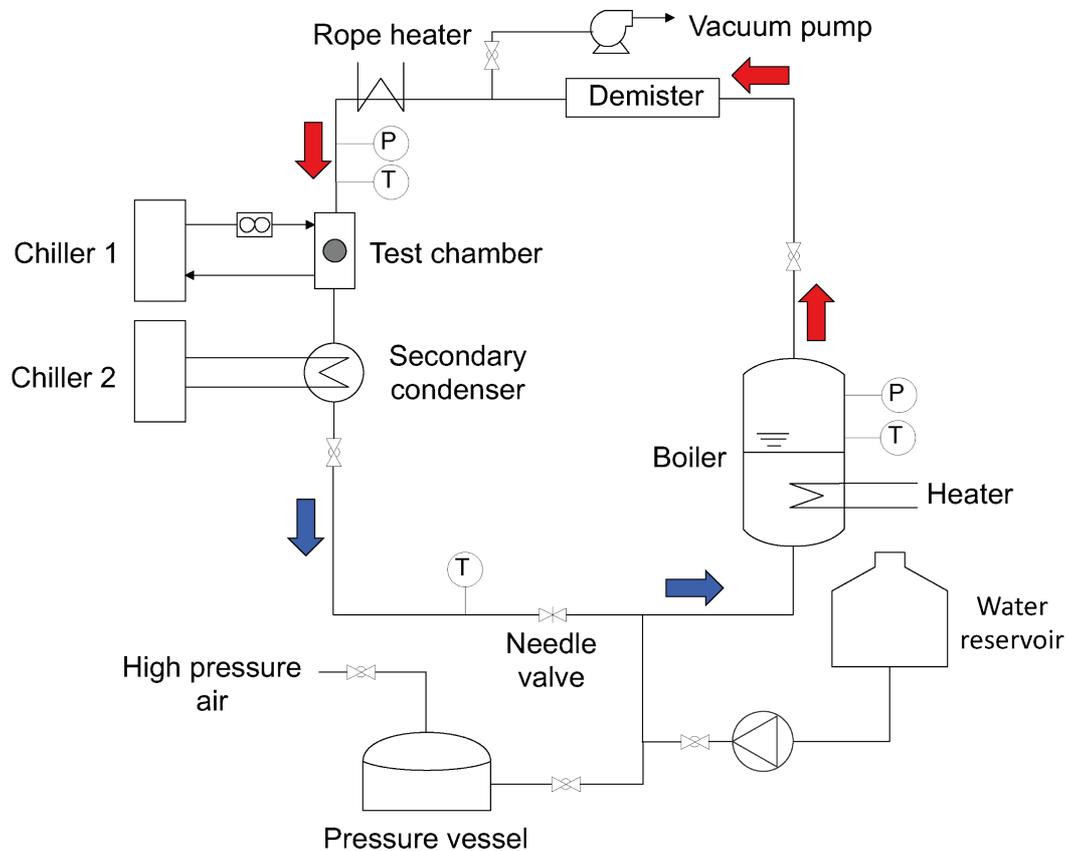

Figure S9.1: Schematic of the overall experimental setup for high-pressure condensation

This section describes the high-pressure flow chamber and experimental procedures.[5] A schematic of the overall setup is shown in Figure S9.1. The high-pressure flow chamber is installed into a loop so that experiments can be run continuously for an extended period. A thermally insulated boiler filled with deionized water is heated to produce steam. The steam flow velocity is regulated by the heating power, which determines the boiling rate. The generated steam flows through a demister to remove liquid droplets carried by the vapor flow. A rope heater is installed along the pipeline to control the steam temperature entering the chamber and ensure dry steam. For all experiments, the steam temperature is kept between 110.7 °C and 111.3 °C. The steam then flows over the test surface, which is mounted on one end of a copper block. The other end of the block is cooled by a recirculating chiller (Chiller 1). 5 thermocouples are placed at precisely known locations along the copper cooler between the two ends to form an array and measure the thermal gradient. Refer to Figure S9.2 for a cross-sectional top view of the flow chamber. One-dimensional heat conduction along the array is ensured by insulating the sides with an air gap. Indium is used as the thermal interface material to enhance thermal contact. The remaining steam not condensed by the test surface then flows out of the chamber and enters the secondary condenser, which is cooled by another recirculating chiller (Chiller 2) and condenses all of the remaining steam into liquid water. The liquid water flows back to the boiler and completes the loop.

The loop pressure is regulated by a pressure vessel and a needle valve. The boiler pressure is also constantly monitored and kept between 1.445 bar and 1.455 bar for all experiments.

At the beginning of each experiment, the chamber is evacuated with a vacuum pump until its pressure reaches below 20 mbar to remove non-condensable gases. Deionized water is then pumped from the reservoir into the setup until a pressure of ~ 1.7 bar is reached. As the system stabilizes, residual gases in the setup rise to the top, where a valve is installed to remove them. Water is pumped into the setup again and the whole process is repeated until no residual gases are expelled. The pressure of the setup is not allowed to drop below atmospheric pressure after the first water pumping exercise in order to avoid inward leakage of air.

Steam is then generated by heating up the boiler and reducing the setup pressure with the pressure vessel by extracting pre-filled pressurized air. Once steam flow is established, a needle valve is used to stabilize it. The minimum and maximum attainable flow velocities are 3 m/s and 9 m/s, at which we report our heat transfer measurements. To achieve different subcooling values, we vary the temperature of the coolant in Chiller 1.

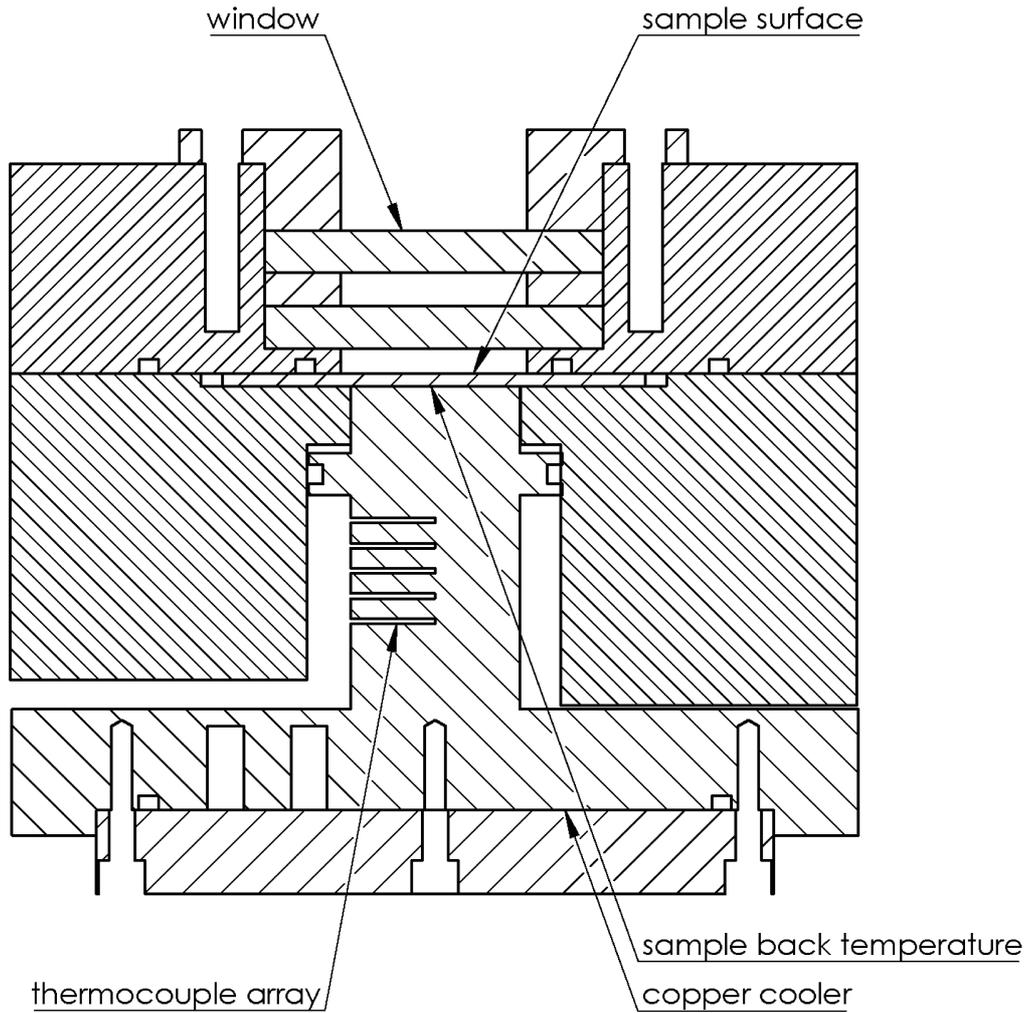

Figure S9.2: Schematic of the high-pressure condensation chamber, top view cross-section.

All sensors are connected to a data acquisition system. The heat flux is computed similarly to the low-pressure flow chamber by linear fitting. A good fit validates the assumption of one-dimensional heat conduction along the thermocouple array.

We estimate the vapor mass flow rate $\dot{m}$ from the heating power of the boiler:

$$\dot{m} = \frac{q_h}{h_{\text{outlet}} - h_{\text{inlet}}}$$

where $q_h$ is the heating power of the boiler, and $h_{\text{outlet}} - h_{\text{inlet}}$ is the difference in enthalpy of the water between the boiler outlet and inlet.

### (b) Computation of heat transfer coefficients

The HTC $h$ is computed similarly to the low-pressure flow chamber, except $A_c \approx A_e$ in the high-pressure flow chamber.

The steam temperature $T_{inf}$ is determined with a thermocouple. An RTD is attached in a slot at the back of our samples and measures the temperature at that location, the sample back temperature $T_b$. It is used to estimate the surface temperature $T_s$:

$$T_s = T_b + q'' \frac{t_{substrate}}{k_{substrate}}$$

where $t_{substrate}$ is the thickness of the sample substrate, and $k_{substrate}$ is the thermal conductivity of the sample substrate. We assume the substrate thickness to be sufficiently thin such that heat conduction is confined to the region between the exposed condensing surface and that in contact with the cooler, without significant diffusion to the sides of the sample used for mounting purposes. The heat flux $q''$, which is based on $A_c$ instead of the full conduction area, is therefore used.

Since only the thickness of the substrate is considered, the estimated surface temperature $T_s$ is effectively the temperature under the coating, if it is present. The computed $T_s$ value is thus lower than the actual one, and the computed subcooling is higher the than actual too. This translates to an underestimation in our HTCs for the high-pressure flow chamber.

The heat transfer values reported are based on measurements at steady states, during which a set of measured quantities have to satisfy a criterion. We set the steady state criterion to be the maximum standard deviation of 120 readings over 2 minutes. The maximum standard deviation allowed is chosen individually for each quantity of the set. If steady state is confirmed, the mean values over the two minutes are accepted as the steady state values.

**(c) Uncertainty propagation**

Uncertainty propagation is computed for measurements in the high-pressure flow chamber.[5]

(i) Uncertainty in subcooling

The uncertainty in steam temperature $\delta_{T_{\text{inf}}}$ is provided by:

$$\delta_{T_{\text{inf}}} = \sqrt{\sigma_{T_{\text{inf}}}^2 + \delta_{T_{\text{inf, cal}}}^2}$$

where $\sigma_{T_{\text{inf}}}$ is the 120-reading standard deviation of the steam temperature and $\delta_{T_{\text{inf, cal}}}$ is the uncertainty from the calibration of the thermocouple.

Similarly, the uncertainty in the sample back temperature $\delta_{T_b}$ is provided by:

$$\delta_{T_b} = \sqrt{\sigma_{T_b}^2 + \delta_{T_b}^2}$$

where $\sigma_{T_b}$ is the 120-reading standard deviation of the sample back temperature and $\delta_{T_b}$ is the uncertainty from the accuracy of the RTD.

The uncertainty in the estimated surface temperature $\delta_{T_s}$ is then propagated:

$$\delta_{T_s} = \sqrt{\delta_{T_b}^2 + \delta_{q''}^2 \left(\frac{t_{\text{substrate}}}{k_{\text{substrate}}}\right)^2}$$

and the uncertainty in subcooling can be computed:

$$\delta_{\Delta T} = \sqrt{\delta_{T_{\text{inf}}}^2 + \delta_{T_s}^2}$$

(ii) Uncertainty in heat flux

The uncertainty in heat flux equals from the uncertainty in the determination of the slope of the temperature gradient. The uncertainty in the linear fit $\delta_{\text{lin\_fit}}$ is calculated as follows, similar to that of the low-pressure flow chamber:

$$\delta_{\text{lin\_fit}} = \delta_{\text{TC}} \sqrt{\frac{N_{\text{TC}}}{N_{\text{TC}} \sum x_{\text{TC}}^2 - (\sum x_{\text{TC}})^2}}$$

where $\delta_{\text{TC}}$ is the uncertainty in the temperature measurement of a thermocouple, $N_{\text{TC}} = 5$ is the total number of thermocouples and $x_{\text{TC}}$ is the location of the thermocouples in the array. We estimate $\delta_{\text{TC}}$ with the expression:

$$\delta_{\text{TC}} = \sqrt{\sigma_{T,\text{max}}^2 + \delta_{\text{cal}}^2 + \delta_{\text{eval, lin\_fit}}^2}$$

where $\sigma_{T,\text{max}}$ is the maximum 120-reading standard deviation from the thermocouples (thus maximum of 5 standard deviation values) during the steady state measurement, $\delta_{\text{cal}}$ is the uncertainty provided by the calibration of the thermocouples, and $\delta_{\text{eval, lin\_fit}}$ is the uncertainty in the evaluation of the linear fit, computed as follows:

$$\delta_{\text{eval, lin\_fit}} = \sqrt{\frac{1}{N_{\text{TC}} - 2} \sum_{i=1}^{N_{\text{TC}}} (T_i - B - A x_i)^2}$$

where $T_i$ is the temperature measured by the $i^{\text{th}}$ thermocouple, and $x_i$ is its location along the array axis.

The uncertainty in heat flux is the uncertainty in the linear fit multiplied by the thermal conductivity of the cooler:

$$\delta_{q''} = k\delta_{\text{lin\_fit}}$$

(iii) Uncertainty in HTC

HTC is obtained from the division of heat flux by subcooling. The uncertainty is propagated in the same way as the low-pressure flow chamber:

$$\delta_h = \sqrt{\left[\frac{1}{(\Delta T)}\delta_{q''}\right]^2 + \left[\frac{-q''}{(\Delta T)^2}\delta_{T_{\text{inf}}}\right]^2 + \left[\frac{q''}{(\Delta T)^2}\delta_{T_s}\right]^2}$$

where $\delta_{T_{\text{inf}}}$ is the uncertainty in the steam temperature, and $\delta_{T_s}$ is the uncertainty in the surface temperature.

The above calculations result in a maximum uncertainty of ~ 11% in HTC and heat flux.

## S10. Shear stress calculation

We compute in this section the shear stress exerted by the ~ 3 m/s steam flow on the tested surfaces in the high-pressure flow chamber (steam at 1.42 bar and 111 °C).

Before the chamber is brought into steady operation, liquid water flows across the surface before it gradually turns into steam. During steady operation, condensation occurs on the surface and condensate droplets of different sizes and weights are formed. The contribution to shear stresses from these 2 factors are not taken into account in the following calculation.

The cross section of the rectangular flow channel is shown in Figure S10.1, where the dimensions $a$ and $b$ are 11 mm and 2 mm respectively, resulting in an aspect ratio of $\phi = 2b/2a = 0.18$ and a hydraulic diameter of $D_H = (4 \times 2a \times 2b)/[2 \times (2a + 2b)] = 6.77$ mm. Assuming a fully developed flow, we then proceed to compute the Reynolds number Re at the aforementioned steam conditions:

$$\text{Re} = \frac{\rho U D_H}{\mu} \approx 1300$$

where $\rho = 0.82$ kg/m³, $U = 3$ m/s and $\mu = 12.62$ µPa s are the density, mean steam velocity, and the dynamic viscosity respectively. The steam properties are obtained with the code CoolProp. We determine the flow to be laminar as $\text{Re} < \text{Re}_{\text{crit}} = 2300$, the critical Reynolds number for turbulent flow in a duct.[6]

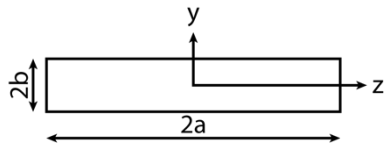

Figure S10.1: Schematic of the cross section of the steam flow channel in the high-pressure flow chamber. Condensing surface is mounted on the wide side at $y = -b$ as steam flows along the x-direction, i.e. perpendicular to the plane of the figure.

The flow profile is then approximated,[7] using the following equation for $\phi \leq 0.5$:

$$u(y,z) = \varepsilon \left[1 - \left(\frac{y}{b}\right)^n\right]\left[1 - \left(\frac{z}{a}\right)^m\right]$$

where

$$\varepsilon = u_m \left(\frac{m+1}{m}\right)\left(\frac{n+1}{n}\right)$$

The parameters are $m = 1.7 + 0.5\phi^{-1.4} = 7.14$ and $n = 2$ respectively. We thus obtain $\varepsilon = 5.13$ m/s. The shear stress exerted on the surface is defined by:

$$\tau_w(z) = \mu \left.\frac{\partial u(y,z)}{\partial y}\right|_{y=-b}$$

Thus

$$\tau_w(z) = \mu\varepsilon \frac{n}{b}\left[1 - \left(\frac{z}{a}\right)^m\right]$$

The maximum shear stress $\tau_{w,\,max}$ is at $z = 0$:

$$\tau_{w,\,max} = \tau_w(z=0) = 64.74 \text{ mPa}$$

We compute the mean shear stress $\bar{\tau}_w$ by integration:

$$\bar{\tau}_w = \frac{1}{2a}\int_{-a}^{a}\tau_w(z)\,dz$$

which gives

$$\bar{\tau}_w = \frac{\mu\varepsilon n}{2ab}\left[2a - \left(\frac{1}{a}\right)^m\left(\frac{1}{m+1}\right)\left(a^{m+1}-(-a^{m+1})\right)\right] = 56.79\text{ mPa}$$

The maximum shear stress translates to an equivalent mass load of 94% of the own mass of the PTFE/CNF coating. This estimate, however, remains highly conservative since the leaching effect of condensate droplets on the coating is excluded from the calculation.

**Videos**

**V1. Condensation on PTFE/CNF in ESEM (MP4)**

Condensation dynamics on PTFE/CNF are visualized in an ESEM. Two different outcomes from droplet coalescence are observed, namely a larger droplet at rest, and droplet jumping out of the field of view.

**V2. Jumping dropwise condensation on PTFE/CNF (MP4)**

Highly frequent jumping droplets from the PTFE/CNF surface observed from the side. Jumping of relatively large droplets with diameter ≈ 0.75 mm is observed.

**V3. Continuous condensation on PTFE/CNF for 72 hours (MP4)**

Condensation behavior on PTFE/CNF in the high-pressure flow chamber at ~ 3 m/s steam flow speed for a continuous period of 72 h. JDWC is sustained as the predominant mode of condensation for 10 h, followed by 50 h of DWC before finally deteriorating to FWC.